\documentclass[aps,prb,reprint,twocolumn,superscriptaddress,amsmath,amsfonts]{revtex4-2}

\usepackage{bm}
\usepackage{amsmath}
\usepackage{amssymb}
\usepackage{mathtools}
\usepackage{mathrsfs}
\usepackage{graphicx}
\usepackage{hyperref}
\hypersetup{
    colorlinks=true,
    linkcolor=MidnightBlue,
    urlcolor=Magenta,
    citecolor=Mulberry
}
\usepackage[dvipsnames]{xcolor}
\usepackage[capitalize]{cleveref}
\usepackage{physics}
\usepackage{dcolumn}

\newcolumntype{d}[1]{D{.}{.}{#1}}

\begin{document}

\title{Statistical Mechanics of Stochastic Quantum Control: $d$-adic R\'enyi Circuits}

\author{Andrew A. Allocca}
\affiliation{Department of Physics and Astronomy, Louisiana State University, Baton Rouge, LA 70803, USA}
\affiliation{Center for Computation and Technology, Louisiana State University, Baton Rouge, LA 70803, USA}
\author{Conner LeMaire}
\affiliation{Department of Physics and Astronomy, Louisiana State University, Baton Rouge, LA 70803, USA}
\author{Thomas Iadecola}
\affiliation{Department of Physics and Astronomy, Iowa State University, Ames, IA 50011, USA}
\affiliation{Ames National Laboratory, Ames, IA 50011, USA}
\author{Justin H. Wilson}
\affiliation{Department of Physics and Astronomy, Louisiana State University, Baton Rouge, LA 70803, USA}
\affiliation{Center for Computation and Technology, Louisiana State University, Baton Rouge, LA 70803, USA}

\date{\today}

\begin{abstract}
  The dynamics of quantum information in many-body systems with large onsite Hilbert space dimension admits an enlightening description in terms of effective statistical mechanics models.
  Motivated by this fact, we reveal a connection between three separate models: the classically chaotic $d$-adic R\'{e}nyi map with stochastic control, a quantum analog of this map for qudits, and a Potts model on a random graph.
  The classical model and its quantum analog share a transition between chaotic and controlled phases, driven by a randomly applied control map that attempts to order the system.
  In the quantum model, the control map necessitates measurements that concurrently drive a phase transition in the entanglement content of the late-time steady state.
  To explore the interplay of the control and entanglement transitions, we derive an effective Potts model from the quantum model and use it to probe information-theoretic quantities that witness both transitions.
  The entanglement transition is found to be in the bond-percolation universality class, consistent with other measurement-induced phase transitions, while the control transition is governed by a classical random walk.
  These two phase transitions merge as a function of model parameters, consistent with behavior observed in previous small-size numerical studies of the quantum model.
\end{abstract}

\maketitle


\section{Introduction}

\begin{figure}
    \centering
    \includegraphics[width=\columnwidth]{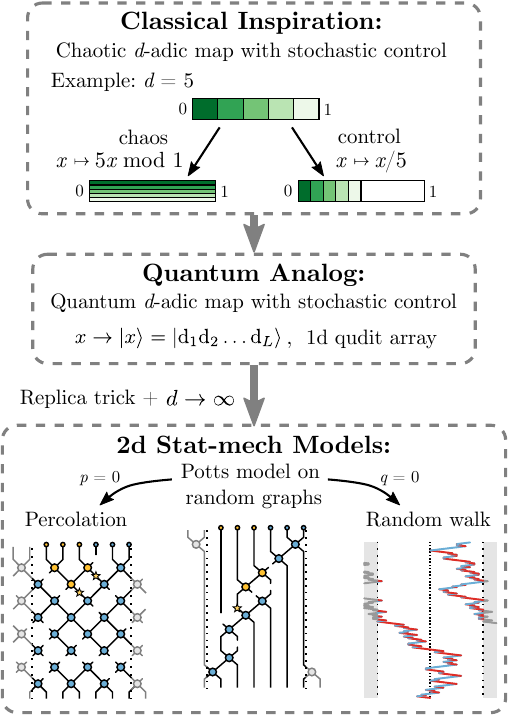}
    \caption{The relationship between models considered in this work we present. 
    The classically chaotic $d$-adic map with stochastic control inspires a quantum analog, from which we then derive a statistical mechanics model via the procedure of Ref.~\cite{Jian2020}.
    Particular limits of this model then exhibit percolation ($p=0$) or random walk ($q=0$) universality.}
    \label{fig:modelflowchart}
\end{figure}

Mappings between quantum systems and classical statistical mechanics models have proved useful for understanding the dynamics of quantum information~\cite{Dennis2002,Nahum17,Zhou19,Vasseur2019,Jian2020,Bao2020,Lopez20,Nahum21,Li21,Barratt22,Nahum23,Majidy23}.
For example, these models have aided in identifying the universal behavior of random quantum circuits~\cite{FisherVijay2023} and measurement-induced phase transitions (MIPTs)~\cite{Li18,Skinner2019,Li19,Chan19,PotterVasseur2022a}.
Simultaneously, the study of adaptive quantum dynamics driven by measurements and unitary feedback operations has been a topic of growing interest~\cite{Iadecola2023,Herasymenko23,Buchhold2022,Ravindranath2022,ODea2022,Friedman2022b,MilekhinPopov2023,PiroliNahum2023,SierantTurkeshi2022,SierantTurkeshi2023,LeMaire2023,Ravindranath23}, with possible applications in state preparation and measurement-based quantum computing~\cite{Briegel2009,Tantivasadakarn22,Lu22,Tantivasadakarn23,Stephen22,Iqbal23,Foss-Feig23}.
In these systems, information gained from measuring a quantum state is used to steer its dynamics, and these control operations can drive a phase transition between an uncontrolled phase, where the system never reaches the target state, and a controlled phase where it always does.
Such ``control transitions'' have a classical antecedent in chaotic maps under \emph{stochastic control}~\cite{Antoniou1996,Antoniou1997,Antoniou1998}, where the dynamics of the chaotic map are randomly interrupted by a control operation and the transition is driven by the frequency of the interruptions.
In quantum systems, a control transition is generically preceded by an entanglement transition driven by the measurements, where the late-time entanglement content of the quantum state abruptly switches from extensive volume-law to subextensive area-law scaling at a critical control rate \footnote{This statement is strictly about control in the context of state preparation; achieving control in the sense of a nonzero order parameter can be achieved before the entanglement transition \cite{SierantPagano2022}.}.
However, in certain circumstances the entanglement and control transitions can coincide.
To understand the interplay between and critical properties of entanglement and control transitions, we construct a statistical mechanics model for a family of adaptive quantum circuits inspired by the $d$-adic R\'enyi map under stochastic control~\cite{Renyi1957}; this is illustrated in \cref{fig:modelflowchart}.

Exact numerical explorations of the entanglement dynamics in systems of qubits are limited to small system sizes or special types of circuits (e.g.,~Clifford~\cite{SierantTurkeshi2022,Zabalo2020} or free-fermion~\cite{Ravindranath23} circuits).
By promoting qubits, with two states per site, to qudits, with onsite dimension $d$, and taking $d\rightarrow\infty$, one can often find a mapping between the entanglement entropy of the system and the energetics of an effective statistical-mechanics model in one higher dimension~\cite{Jian2020}. 
The power of these mappings lies both in enabling scalable simulations and in providing classical intuition that can enlighten the analysis of the original quantum model at $d=2$.
Our statistical mechanics model adapts tools originally developed in Ref.~\cite{Jian2020} for brickwork circuits of Haar random unitaries in the presence of measurements and takes the form of a Potts model on a random graph. 
This model, based on a stochastic quantum map that is itself inspired by a chaotic classical map (see~\cref{fig:modelflowchart}), represents one of the central results of our work.
We will use it to disentangle and provide intuitive explanations for two critical behaviors, stemming from percolation and random-walk physics, that have been observed in other systems.

\begin{figure}
    \centering
    \includegraphics[width=\columnwidth]{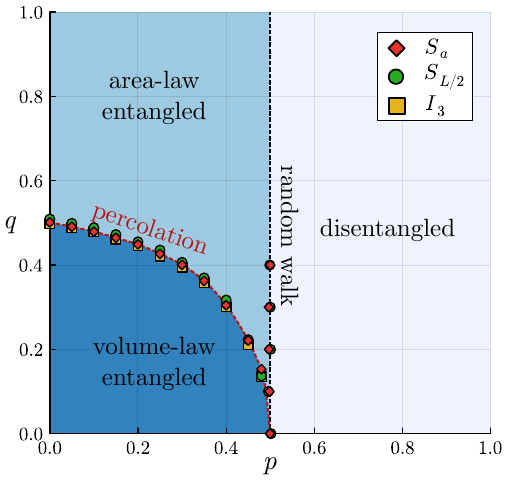}
    \caption{The two-dimensional phase diagram of the stochastic circuit with additional random measurements.
    The parameter $p$ is the probability to apply the control map at each time step, and $q$ is the probability to measure the sites acted on by the unitary after each chaotic step. 
    We determine phase boundary labeled ``percolation'' with three quantities, the ancilla entropy $S_a$, the half-cut entropy $S_{L/2}$, and the tripartite mutual information $I_3$, and confirm the vertical phase boundary labeled ``random walk'' using $S_a$ and $S_{L/2}$. 
    All methods are in good agreement.}
    \label{fig:phasediagram}
\end{figure}

To study the interplay of these two criticalities, the quantum model from which the Potts model is derived features two types of measurements.
One occurs during the application of the control map, which happens with probability $p$ at each time step.
If the control map is not applied, a chaotic, entangling map based on the $d$-adic R\'enyi map is applied instead.
Following application of the chaotic map, we allow additional measurements to occur with a probability $q$.
This leads to a phase diagram, shown in \cref{fig:phasediagram}, with three phases: a volume-law phase where the system remains highly entangled at late times, an area-law phase driven by the measurements with probability $q$, and a disentangled phase driven by the control protocol with probability $p$.
This phase diagram, and associated information about critical properties, is obtained by mapping certain information-theoretic quantities that witness the transitions (e.g.,~entanglement entropies, mutual informations, and purification measures) onto correlation functions in the Potts model.
The transition between the volume- and area-law phases is in the bond-percolation universality class consistent with the ``standard'' MIPT.
The transition between the area-law and disentangled phases, described by a random walk, is in the same universality class as the classical control transition and was also previously observed in direct simulations of the quantum model~\cite{Iadecola2023,LeMaire2023}.
The two transitions coalesce at $q=0$, consistent with predictions from a finite-size scaling analysis in Ref.~\cite{Iadecola2023}.

The paper is organized as follows. 
In \cref{sec:Renyimodel}, we describe the $d$-adic R\'enyi map and its quantum analog that forms the basis of our study. 
We further introduce how adding measurements with probability $q$ will allow us to split the control and entanglement transitions.
In \cref{sec:statmechmodel}, we describe the derivation of the Potts model which describes the information content of the quantum $d$-adic map as well as the observables we measure (in both information-theoretic terms and their analogs in the statistical model).
Following this, in \cref{sec:results}, we uncover the phase diagram and critical properties and discuss their implications for the quantum system.
In \cref{sec:conclusion}, we conclude with an outlook and open questions.

\section{$d$-adic R\'enyi Map Under Stochastic Control: Classical Model and Quantum Analog} 
\label{sec:Renyimodel}

We begin by describing the $d$-adic R\'enyi map under stochastic control and the analogous quantum circuit.
The $d$-adic R\'enyi map~\cite{Renyi1957} acts on a real number $x\in[0,1)$ as follows:
\begin{equation}\label{eq:d-adic}
    x\mapsto d\,x \mod 1.
\end{equation}
The dynamics of any irrational number $x_0$ under this map is chaotic in the sense that its orbit under the map, $x_0\to x_1\to x_2 \to \dots$, uniformly fills the unit interval.
However, rational initial conditions generically undergo periodic orbits.
For example, for $d=2$ the initial condition $x_0=1/3$ undergoes a period-2 orbit $1/3 \to 2/3 \to 1/3 \to \dots$.
In addition, for any $d$, the origin $x=0$ is a fixed point.
These regular dynamical trajectories and fixed points are unstable in the sense that any rational $x_0$ has an irrational number arbitrarily close, so that any infinitesimal displacement of the initial condition will generically lead to chaotic dynamics.

It was demonstrated in Ref.~\cite{Antoniou1996} that these unstable orbits can be transmuted into global attractors by a stochastic control protocol.
The protocol defines a modified discrete-time dynamics wherein, at each time step, a control operation is applied with probability $p$, and otherwise the chaotic map \cref{eq:d-adic} is applied.
The control map is designed to have fixed points corresponding to each point on the orbit it targets.
In the simplest case of control onto the fixed point $x=0$, the control map is given by the contraction
\begin{equation} \label{eq:controlmap}
    x\mapsto (1-a) x,
\end{equation}
where $0\leq a\leq 1$ parameterizes the strength of the control.
At fixed $0<a<1$ the dynamics of the coupled chaotic and control maps exhibits a phase transition at~\cite{Antoniou1998}
\begin{align}
p_c = \frac{\ln(1-a)}{\ln(1-a)-\ln(d)}.
\end{align}
For $p>p_c$, any initial condition eventually reaches the orbit, whereas for $p<p_c$ it does not.

To define a quantum analog of this model, we need to replace the classical state $x$ with a quantum computational basis (CB) state $\ket{x}$. 
To do this, we first expand $x$ in a base-$d$ fraction,
\begin{align}
    x \equiv 0.d_1 d_2 \dots =  \sum^{\infty}_{k=1} \frac{d_k}{d^k},
\end{align}
where the individual ``dits" $d_k$ may take values $0,1,\dots,d-1$.
The chaotic map \cref{eq:d-adic} acts by ``shifting the decimal point'' to the right in the dit string representation and throwing away the most significant dit, $0.d_1d_2\dots \mapsto 0.d_2d_3\cdots$.
Upon setting $a=(d-1)/d$ the control map \cref{eq:controlmap} becomes division by $d$ and takes a similarly simple form: a leftward shift of the decimal point, $0.d_1d_2\dots \mapsto 0.0d_1d_2\cdots$.
For this value of $a$, the control transition occurs at $p_c=1/2$, where the decimal point effectively undergoes an unbiased random walk.

This is an exact rewriting of the classical model in a discretized form and a quantum model is naturally acquired by promoting dits to $d$-state qudits. 
In order for this quantum model to remain tractable, however, we need a finite dimensional Hilbert space, which we acquire by first truncating the classical system to $L$ dits.
This change necessitates modifying details of the dynamics.
First, instead of discarding the leftmost (rightmost) dit with each chaotic (control) map, we cycle it to the end (beginning) of the string. 
This can be done with either an ascending or descending staircase of SWAP operations, or can be framed as leaving the dits in place and moving the decimal point through a periodic system.
We use the latter picture. 
After moving the decimal point, the control map then sets the dit to the right of decimal point to zero, and the chaotic map randomizes the two dits to the left of the decimal.
The randomization is introduced due to the fact that truncation has removed the possibility of chaos---indeed, without any randomization, such truncated numbers are rational and therefore will not undergo chaotic dynamics under a simple cyclic translation.
This randomization operation can be written
\begin{equation}
    \cdots d_i \,{}_\bullet\, d_{i+1} d_{i+2} \cdots \mapsto \cdots \tilde{d}_i \tilde{d}_{i+1} \,{}_\bullet\, d_{i+2} \cdots,
\end{equation}
and simulates the distribution of digits in almost all irrational numbers~\cite{emile_borel_les_1909, Iadecola2023}; this allows for a faithful representation of the $d$-adic map's chaotic properties while sacrificing determinism~\cite{LeMaire2023}.

Exchanging dits for qudits, we now have $L$ quantum degrees of freedom and a single classical piece of information, the position of the decimal point.
The movement of the decimal point determines the structure of the quantum circuit that evolves the qudits in time as we now describe. 
To obtain quantum dynamics in the system, we promote the random but classical scrambling function to a random unitary transformation: 
\begin{equation}\label{eq:quantum_d-adic}
    \ket{\cdots d_i \,{}_\bullet\, d_{i+1} d_{i+2} \cdots}\mapsto U_{i,i+1}\ket{\cdots d_i d_{i+1} \,{}_\bullet \, d_{i+2} \cdots}.
\end{equation}
First, the decimal point is translated to the right, then a random two-qudit unitary operator $U$, drawn uniformly from the Haar measure on $U(d^2)$, is applied to the two sites immediately to the left of the decimal point, so that $U_{i,i+1} = \openone_{d^{i-1}} \otimes U \otimes \openone_{d^{L-i-1}}$ ($\openone_{N}$ is an $N\times N$ identity matrix) and its periodic extension for $U_{L,1}$. 
This generates superposition and entanglement of the computational basis states, leading to quantum chaotic behavior~\cite{Iadecola2023}. 
In this case, the model becomes challenging to simulate owing to the exponential growth of the $d^L$-dimensional Hilbert space.
A more modest polynomial scaling can be achieved by restricting $U$ to be a random two-qudit Clifford gate~\cite{Mari12,Gross2006} at the expense of modifying the physics in a manner discussed at the end of this section.
The control map is implemented via the following nonunitary transformation:
\begin{align}
    \ket{\cdots d_{i-1} d_i \,{}_\bullet\, d_{i+1} \cdots}\mapsto \ket{\cdots d_{i-1} \,{}_\bullet\, 0_i d_{i+1} \cdots}.
\end{align}
This \textit{reset operation} can be implemented as (i) a projective measurement on the qudit to the left of the decimal point, (ii) a unitary rotation on that qudit, conditional on the measurement outcome, to the $\ket{0}$ state, and (iii) a translation of the decimal point to the left.
Note that where the chaotic and control maps are applied is determined by the location of the decimal point, which moves at each time step based on which map is randomly chosen according to the value of $p$. 
Regardless of the nature of the scrambling operation $U$ (permutation, Haar-random, or Clifford), the location of the classical control transition at $p_c=1/2$ is unchanged by the modifications to the original classical model because it is determined by the motion of this classical piece of information, a fact verified in Refs.~\cite{Iadecola2023,LeMaire2023}.

In addition to the control transition inherited from the classical model, the quantum model exhibits a volume-to-area-law entanglement transition driven by the measurements performed as part of the control map. 
In Ref.~\cite{Iadecola2023}, which considered $d=2$ and Haar-random gates, the control and entanglement transitions were found to coincide to within the precision set by the small-size classical numerical methods used. 
Nevertheless, it was argued there that the transitions should coincide owing to the fact that entanglement spreads throughout the system only via the translations applied as part of the chaotic map.
Reference \cite{LeMaire2023} reached larger system sizes by replacing the Haar-random gates with Clifford gates, and it was found that the entanglement and control transitions separate, with the former preceding the latter as expected for generic measurement-and-feedback-driven transitions~\cite{ODea2022,Ravindranath2022}.
This was attributed to the fact that Clifford gates have a finite probability of locally disentangling a stabilizer state.
Moreover, the criticality of the entanglement transition was broadly consistent with that of the MIPT in Clifford circuits.
Note that in this setting of two separate transitions, the control transition also has entanglement signatures: the entanglement changes from area-law to fully disentangled (i.e., from finite to zero) and exhibits distinct critical scaling properties associated with the random walk.

\begin{figure}
    \centering
    \includegraphics[width=0.8\columnwidth]{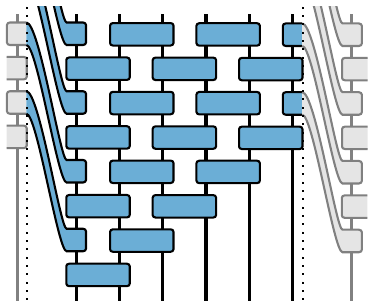}
    \caption{A schematic showing the twisted boundary conditions generated by our protocol for $p=0$. No measurements are shown here, but would be performed with probability $q$ after each gate.}
    \label{fig:twisted}
\end{figure}

To provide an independent perspective on this observation, and to test whether the entanglement and control transitions can be shown numerically to coincide in larger systems, we add one final ingredient to the $d$-adic R\'enyi circuit described above.
Each time the chaotic map \cref{eq:quantum_d-adic} is applied (which happens with probability $1-p$ at each time step), we allow a computational basis measurements to occur with probability $q$ on the two qudits acted upon by the unitary gate. 
At $p=0$, the circuit becomes a brickwork lattice of two-qudit gates, but with the boundary between the 1st and $L$th qubits twisted by $L-2$ time steps, demonstrated in \cref{fig:twisted}. 
In this limit, we expect to recover the physics of the MIPT, where a volume-to-area-law entanglement transition in the bond-percolation universality class occurs at the percolation threshold $q=1/2$.
In the opposite limit, where $q=0$, we should recover the picture suggested in Ref.~\cite{Iadecola2023} where the entanglement and control transitions coincide.
We expect the two transitions to separate when $p$ and $q$ are both finite, and an example random circuit for generic $p$ and $q$ is shown in \cref{fig:statmechschematic}(d).
However, the control transition line will remain fixed at $p=1/2$ for all $q$, with the entanglement transition line interpolating from $(p,q) = (1/2,0)$ to $(0,1/2)$.
The statistical mechanics model we derive in the next section yields a phase diagram, \cref{fig:phasediagram}, that is consistent with this intuition.

\section{Statistical Mechanics Model and Observables} 
\label{sec:statmechmodel}

We follow the general recipe of Ref.~\cite{Jian2020} to obtain the free energy of a 2+0d classical model starting from the R\'enyi entropies of the 1+1d qudit circuits generated by the stochastic process described above.
The $n$th R\'enyi entropy is
\begin{equation}
    S_{n}(A) = \frac{1}{1-n}\log\Tr_A\left(\rho_A^n\right), 
\end{equation}
written here in terms of the reduced density matrix for a subsystem $A$ obtained by a partial trace of the full density matrix over the degrees of freedom in the complement of $A$, $\rho_A = \Tr_{\bar{A}} \rho$.
For this procedure, however, it is necessary to encode this partial trace in terms of a permutation operator $\mathcal{P}_{n,A}$ acting on the $n$ copies of the full density matrix: in subsystem $A$ this operator permutes between these $n$ copies sequentially, and in subsystem $\bar{A}$ it acts as the identity, resulting in a trace of the degrees of freedom on these sites. 
The R\'enyi entropy rewritten in terms of this operator is then averaged over all quantum trajectories of the circuit $C$ using a replica trick, and finally related to a partition function,
\begin{equation}
    \bar{S}_{n}(A) = \lim_{m\to0}\frac{\mathcal{Z}_A - \mathcal{Z}_0}{m(1-n)},
\end{equation}
where $m$ indexes the artificial replicas and
\begin{gather}
    \mathcal{Z}_A = \mathbb{E}_C\, \Tr\left[(C\rho_0 C^\dagger)^{\otimes Q} \mathcal{P}_{n,A}^{\otimes m}\right],\\
    \mathcal{Z}_\emptyset = \mathbb{E}_C\, \Tr\left(C\rho_0C^\dagger\right)^{\otimes Q}
\end{gather}
with $Q=mn+1$, are the partition functions for the system with and without the partial trace---subtracting $\mathcal{Z}_\emptyset$ removes all effects not related to the choice of $A$.
The density matrix of the initial state $\ket{\psi_0}$ is $\rho_0 = \dyad{\psi_0}$, and $C$ represents the entire quantum circuit, here the stochastic product of projective measurements and unitary gates as described in \cref{sec:Renyimodel}.
In the replica limit $m\to0$ we have $\mathcal{Z}_A \to \mathcal{Z}_\emptyset$, so we can approximate $\mathcal{Z}_A - \mathcal{Z}_\emptyset \approx \log\mathcal{Z}_A-\log\mathcal{Z}_\emptyset \equiv -(F_A - F_\emptyset)$, defining ``free energies'' $F_A$ and $F_\emptyset$ from partition functions in the usual way.
Note that the index $n$, specifying which R\'enyi entropy we are calculating, only enters this expression multiplying the replica index $m$, which is taken to $0$.
As a result, the same result is obtained for all choices for $n$, equivalent to the case of $n=0$, i.e.~the Hartley entropy.
We will therefore drop the index moving forward. 

This procedure maps a circuit onto a $Q!$-state Potts model. 
Unitary gates in the original circuit become sites in a lattice where the variables of this Potts model reside, those being elements of the permutation group acting on the $Q$ total replicas of the system, $\sigma\in \mathcal{S}_Q$.
Measurements in the circuit model remove the Potts interaction between neighboring sites.
The circuit time becomes a second spatial dimension, and the partial trace at the final time imposes boundary conditions on the Potts variables within $A$ and $\bar{A}$: $\sigma=e$ within $\bar{A}$ (the identity permutation), and $\sigma = (12\cdots n)^{\otimes m}$ within $A$. 
The energy of this model satisfying the boundary condition, $F_A-F_\emptyset$, is thus directly related to the averaged entanglement entropy of the circuit.
Note, however, that in this procedure the state of the qudits themselves plays no role and is not preserved, so there can be no sense of ``control'' in the Potts model---it \emph{only} contains information about the entropy.
The control operation in our protocol is therefore realized as just a projective measurement followed by a translation of the decimal point, with the measurement-conditioned single-qudit unitary having no effect.

\begin{figure*}
    \centering
    \includegraphics[width=0.8\textwidth]{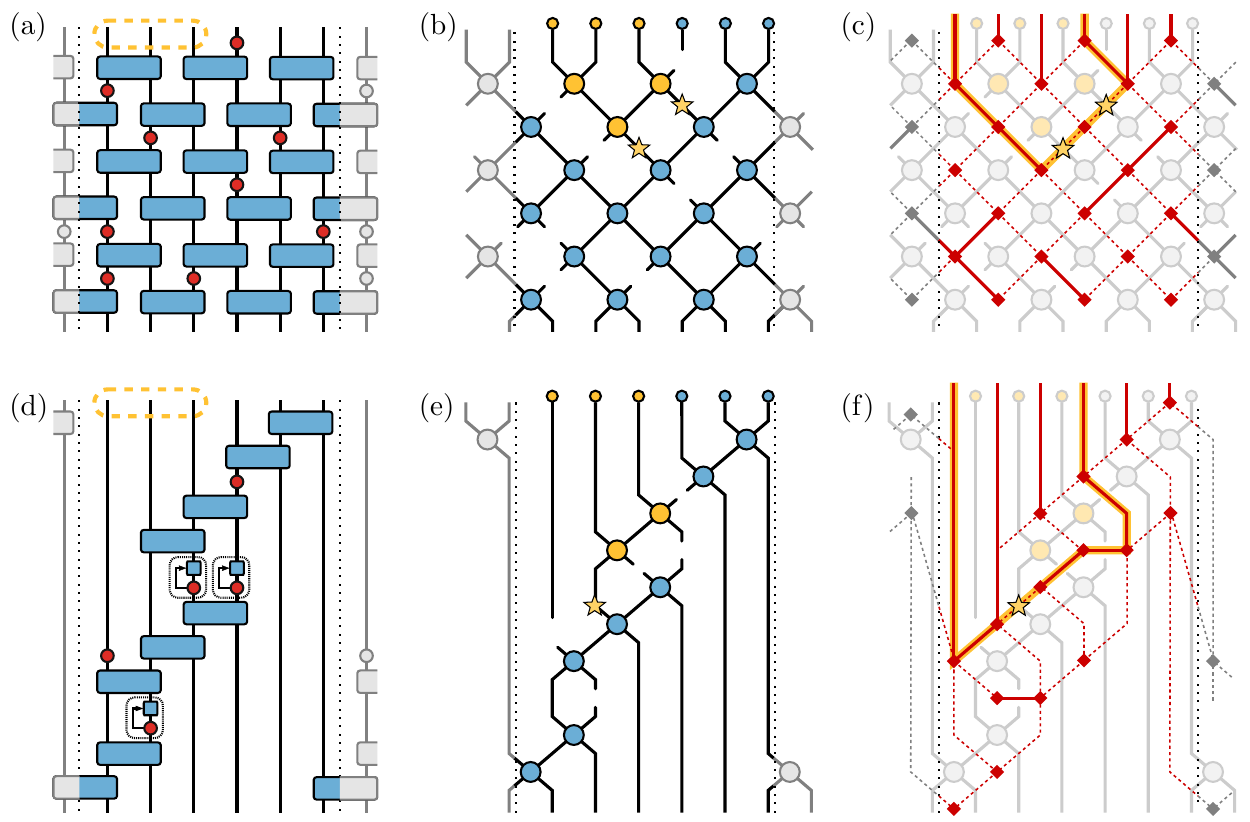}
    \caption{Schematic diagrams showing the relation between quantum circuits, Potts models, and the minimal cut framework for determining entanglement entropies.
    (a) An example brickwork circuit composed of 2-site unitaries (blue rectangles) with random projective measurements (red circles). Subsystem A is composed of the 3 sites at the final time circled by the dashed oval.  
    (b) The tilted square lattice for the $Q!$-state Potts model obtained from (a). Each site (large circles) corresponds to a unitary in the circuit and are where the $Q!$-state ``spin'' variables of the model reside. Each measurement in the circuit yields a cut bond between these sites, i.e.\ no coupling between the respective variables. The boundary conditions in $A$ and $\bar{A}$ are encoded by the small circles at the final time, and determine the state on each site (indicated by color, blue or yellow) that minimizes the free energy $F_A$. The bonds marked with stars are a minimal-length domain wall in this system, and cutting these bonds disconnects the domains containing $A$ and $\bar{A}$---a minimal cut. 
    (c) The dual graph representation of Potts model lattice in (b), allowing for simple determination of the minimal cut via path length minimization. Vertices are shown as red diamonds, length-0 edges as thick red lines, and length-1 edges as dashed red lines. A path corresponding to the domain walls in (b) is highlighted in yellow. 
    (d) An example stochastic circuit. As in (a) subsystem $A$ is circled. The reset operation of the control steps are also shown, feeding measurement outcomes into the dynamics of the circuit. 
    (e) The irregular lattice for the $Q!$-state Potts model corresponding to the circuit (d). The only aspect of the control steps relevant for obtaining this lattice are the measurements.
    (f) The dual graph representation of the Potts model in (e). In this case the shortest path has length 1.}
    \label{fig:statmechschematic}
\end{figure*}

The model derived in Ref.~\cite{Jian2020} for the regular brickwork circuit of gates and random measurements yields a regular tilted square lattice.
An example circuit and its resulting lattice are shown in \cref{fig:statmechschematic}(a) and (b). 
Because measurements remove the interaction between neighboring lattice sites, for $Q\to1$ the Potts model reduces to a picture of bond percolation, so phases of the system, their energies, and therefore the entanglement phases, are simple to deduce.
For small measurement rates few bonds are cut, so a typical realization contains a cluster of connected bonds spanning between $A$ and $\bar{A}$, each pinned to different states.
The energy of the system is then determined by the length of the shortest domain wall separating domains in these different states, which is be proportional to the size of $A$ or $\bar{A}$, whichever is smaller.
The entropy is therefore proportional to the volume of the subsystem.
For high measurement rates many bonds are cut, leaving only small clusters of connected bonds.
Sites in $A$ and $\bar{A}$ are only be connected by small clusters at their boundary in typical realizations, so the domain wall between them has length determined by the size of this boundary.
The entropy is thus proportional to the surface area of the subsystem.
Bond percolation in two dimensions has a phase transition when bonds are cut with probability $1/2$, so this is also the critical measurement rate for this model. 
The length of the domain wall between different orderings can alternatively conceived of as the minimum number of additional bonds that must be cut in order to completely separate these phases imposed by $A$ and $\bar{A}$.
For this reason the length of the domain wall is also referred to as the ``minimal cut.'' 

This domain wall/minimal cut picture can alternatively be phrased in terms of a dual graph: in the Potts model lattice we identify faces as regions bounded by bonds (both cut and uncut), and for each of these faces we draw a vertex. 
Two vertices in the dual graph are connected by an edge of length $0$ if they are separated by a cut bond (measured qudit) in the original graph, and by an edge of length $1$ if they are separated by an uncut bond (unmeasured qudit).
At the final time boundary of the circuit there are $L$ vertices corresponding to the faces between the $L$ qudit lines.
Choosing two of these vertices to bound our complementary regions $A$ and $\bar{A}$, the entanglement entropy is then the smallest length among all possible paths between these vertices; each uncut bond contributes length $1$, so this path minimization counts the minimal number of bonds needing to be cut. 
An example of this procedure is shown in \cref{fig:statmechschematic}(c).
We will use this conceptualization of minimal cuts as minimal path lengths in the dual graph in our further analysis. 

Because our protocol is inherently stochastic, the Potts model resulting from the above procedure for arbitrary $p$ and $q$ is generically defined on an irregular lattice and cannot be analytically analyzed as directly as the case of the regular brickwork circuit.
However, many of the same general results carry through:
we obtain precisely the same mapping from unitary gates to sites in a lattice hosting $Q!$-valued degrees of freedom with measurements cutting the bonds between them, and we can analyze the entanglement entropy as the minimal number of additional bonds that must be cut in order to separate the domains containing subsystems $A$ and $\bar{A}$.
We can also similarly construct the dual graph from which this minimal cut value, and therefore the entanglement entropy, is ascertained via path minimization.
In \cref{fig:statmechschematic}(d)-(f) we show an example stochastic circuit, the related Potts model lattice, and the associated dual graph. 

\subsection{Minimal cut model} 
\label{sec:minimalcutmodel}

By analyzing how distances in the dual graph formulation of our stochastic circuit model are updated as we apply entangling or control steps, we obtain an exact procedure to update the minimal cut for any choice of contiguous subsystem $A$ as the circuit is generated.
We use an $L\times L$ matrix $\hat{d}$ to track the minimal path lengths between the $L$ different possible final-time vertices bounding our subsystem $A$; the element $d_{ij}$ is the distance through the dual graph between vertex $i$ and $j$, so $\hat{d}$ is symmetric with $0$s along the diagonal. 
In each simulation of the circuit we initialize this matrix with some initial condition on these distances encoding the entanglement of the initial state (e.g. $d_{ij} = 0$ for all $i,j$ is a product state), and at each time step this matrix is updated according to whether we perform a chaotic or control step. 
We will label the position of the decimal point and the vertices of the dual graph by the qudit immediately to their left. 

Measuring qudit $i$ cuts the corresponding bond in the Potts model and sets the length of the link between the vertices at $i$ and $i-1$ to zero, $d_{i,i-1} \to 0$.
Consequently, the distance from any other vertex $j$ to $i$ and $i-1$ becomes the minimum of the distances between $j$ and either $i$ or $i-1$ before the map is applied.
We also need to enforce the triangle inequality between all other pairs of vertices $j, k$; since the distances from $j$ and $k$ to $i$ or $i-1$ have possibly decreased, the path from $j$ to $k$ passing through $i$ or $i-1$ may be the new minimal path.
Altogether, we have that the effect of a measurement at $i$ is
\begin{equation} \label{eq:measurement}
\begin{gathered}
    d_{ij} \xrightarrow{M_i} d'_{ij} = \text{min}\left(d_{ij},d_{i-1,j}\right) \\
    d_{i-1,j} \xrightarrow{M_i} d'_{i-1,j} = \text{min}\left(d_{ij},d_{i-1,j}\right) \\
    d_{jk} \xrightarrow{M_i} d'_{jk} = \text{min}\left(d_{jk}, d'_{ij}+d'_{ik}\right) \quad \forall j,k\neq i,i-1,
\end{gathered}
\end{equation}
where $\hat{d}$ is kept symmetric at all points.
As noted previously, since the model does not track the state of the qudits in the system, the control map is simply a measurement implemented in this way followed by a shift of the decimal point. 

Applying the chaotic map first moves the decimal point to the right from $i$ to $i+1$, then applies a random unitary gate to qudits $i$ and $i+1$.
This creates a new ``face'' in the Potts model lattice, corresponding to a new vertex at position $i$ which is only connected to the vertices at $i+1$ and $i-1$.
Therefore, only paths with one endpoint at $i$ are affected, and the elements of $\hat{d}$ are updated due to this unitary gate as
\begin{equation} \label{eq:chaotic}
    d_{ij} \xrightarrow{U_i} d'_{ij} = 1+\mathrm{min}\left(d_{i+1,j},d_{i-1,j}\right) \quad \forall j\neq i.
\end{equation} 
Since $d_{ij} \leq d_{ik} + d_{kj}$ for all $k$ by the triangle inequality we can be assured that this update is the minimal new distance between $i$ and $j$.
We also need to enforce that the maximum value of $d_{ij}$ is the actual physical distance between $i$ and $j$.
For instance, the distance between vertices $1$ and $4$ is never greater than $3$ for $L\geq6$---at worst we can cut through the $3$ qudit lines between the corresponding faces.
We then allow for two measurements on sites $i$ and $i+1$, each with probability $q$, as given in \cref{eq:measurement}.  

After stochastically running these dynamics, the elements of $\hat{d}$ record the minimal cut corresponding to \emph{any} choice of connected final-time subsystem $A$.
A natural choice for $A$ is the leading $L/2$ qubits, using the location of the decimal point to set one boundary, giving the half-cut entropy $S_{L/2}$.
The other values in $\hat{d}$ are useful in defining other measures of entanglement such as the bipartite or tripartite mutual information, $I_2$ and $I_3$.

\subsection{Ancilla model} 
\label{sec:ancillamodel}

Another useful way to analyze the entanglement properties of a system is in terms of a purification transition~\cite{Gullans2020a,Gullans2020b}.
Consider maximally entangling an $L$ qubit system at the initial time with an ancilla qubit, which is not acted upon by the subsequent dynamics. 
In the volume-law phase the system remains in a mixed state for exponentially long time, while in area-law or disentangled phases it purifies on much shorter time scales.
Therefore, by choosing an appropriate depth for the circuit, the rank of the reduced density matrix of the ancilla at the final time reveals the entanglement phase the system. 

\begin{figure}
    \centering
    \includegraphics[width=\columnwidth]{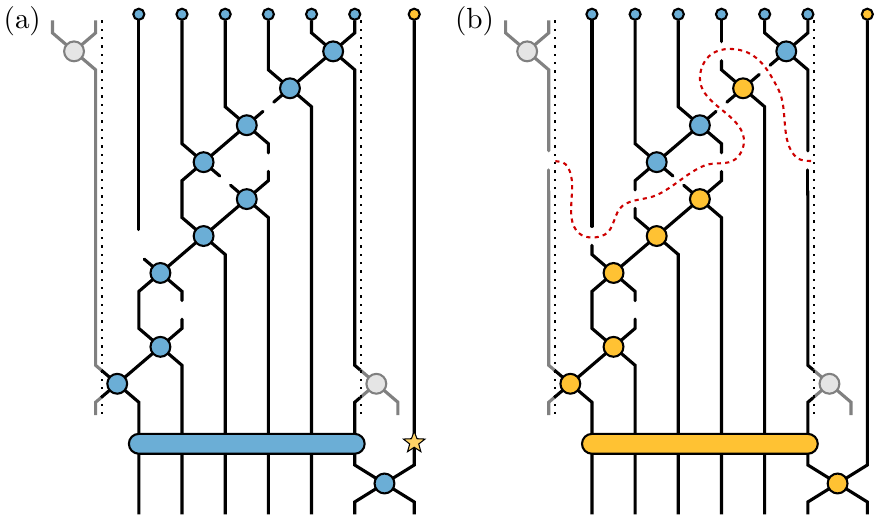}
    \caption{Example Potts model lattices corresponding to stochastically generated circuits including an ancilla qudit (far right line) that is maximally entangled at the initial time. (a) A realization with $S_a = 1$. The domain containing the ancilla can be disconnected from the rest of the system by a single cut, indicated with a star. (b) A realization with $S_a = 0$---the same sequence of chaotic and control steps as in (a), but with more measurements following unitary gates. There are enough cut bonds to disconnect the ancilla domain from the domain containing the $L$ qudits at final time; the dashed red line shows the separation.}
    \label{fig:ancillaschematic}
\end{figure}

Applying the procedure outlined above but now with an ancilla qudit maximally entangled at the initial time, we again obtain a $Q!$-state Potts model that encodes this ancilla entropy as the energy of a domain wall in the system.
The subsystem $A$ is now the ancilla itself and the $L$ qudits of the system after stochastic evolution comprise $\bar{A}$. 
Because all degrees of freedom are maximally entangled at the initial time---all sites in the Potts model lattice are connected---the domain wall separating $A$ and $\bar{A}$ is either (i) a space-like curve through the $L$-qudit system separating these qudits at final-time from the initial state, or (ii) a cut through the ancilla itself.
The latter of these sets the maximum value for the minimal cut to be $1$, and the only case where the former gives a smaller minimal cut is if a \emph{trivial} domain wall exists requiring $0$ cuts, meaning that the final-time state of the $L$-qudit system is entirely uncorrelated with its maximally-entangled initial state---the system has lost its memory of the initial state.
These two cases are demonstrated in \cref{fig:ancillaschematic}. 

Calculating the ancilla entropy is thus equivalent to determining if there exists a path through the Potts model lattice from the ancilla, i.e. the initial-time boundary where all qudits and the ancilla are connected, to at least one qudit at the final time of the circuit. 
Just as we determined the minimal cut number above via updating the matrix $\hat{d}$ according to \cref{eq:measurement,eq:chaotic}, we can determine if such a path exists now using two pieces of classical information that are updated as the circuit is constructed. 
First is an $L$-bit list $a$ which tracks the connection of the ancilla through the Potts model lattice to each qudit at the present time step: 
if $a_i = 1$ then there is a path along uncut bonds from the ancilla to qubit $i$, and if $a_i = 0$ then there is not.
Second is a symmetric $L\times L$ matrix $\hat{c}$ tracking the connectivity of the lattice at the present time step. 
If $c_{ij} = 1$ then there is a path along uncut bonds through the circuit between qudits $i$ and $j$, and if $c_{ij} = 0$ then there is not. 
Initially, the system is in a maximally entangled state, so all entries of $a$ and $\hat{c}$ are set to $1$.

Whenever we measure a site $i$, it is disconnected from every other qudit and from the ancilla, so
\begin{equation} \label{eq:anccontrol}
\begin{gathered}
    c_{ij} \xrightarrow{M_i} \delta_{ij}\\
    a_i \xrightarrow{M_i} 0.
\end{gathered}
\end{equation}
When a unitary gate is applied between $i$ and $i+1$ then they become connected, so every site $j$ connected to $i$ also becomes connected to $i+1$, and vice versa.
Additionally, all sites $j$ connected to $i$ are now also connected to all sites $k$ connected to $i+1$. 
Altogether the action of a unitary gate is
\begin{equation} \label{eq:ancgate}
\begin{aligned}
    c_{i,i+1} &\xrightarrow{U_i} 1\\
    c_{ij},c_{i+1,j} &\xrightarrow{U_i} c'_{ij},c'_{i+1,j} = c_{ij}\,\lor\, c_{i+1,j} \quad \forall\, j\neq i,i+1\\
    c_{jk} &\xrightarrow{U_i} c_{jk}\,\lor\,(c'_{ij}\,\land\,c'_{ik}) \quad \forall\, j,k \neq i,i+1 \\
    a_j &\xrightarrow{U_i} 
    \begin{cases} 
        0, & \sum_k c_{jk}a_k = 0,\\
        1, & \sum_k c_{jk}a_k \neq 0,
    \end{cases}\\
\end{aligned}
\end{equation}
where $\lor$ is the bitwise OR operation and $\land$ is the bitwise AND operation. 
Note that in both \cref{eq:anccontrol,eq:ancgate} we must also ensure that $\hat{c}$ remains symmetric. 
As for the algorithm tracking the minimal path lengths above, the control map is a measurement followed by a shift of the decimal point, and a chaotic step entails is a shift of the decimal point, a unitary gate, and two probabilistic additional measurements. 
For a given realization of the stochastic circuit, the ancilla entanglement entropy $S_a$ is $0$ if all the entries of $a$ are $0$, and $1$ otherwise. 

\subsection{Observables} 
\label{sec:observables}

\begin{table*}[]
    \centering
    \begin{tabular}{c|c|c}
    \hline \hline
         Notation & Information theoretic quantity & Percolation quantity   \\ \hline
         $S(A)$ & Entanglement entropy of region $A$ & Minimal cut connecting points on the boundary of $A$ \\
         $S_a$ & Ancilla entropy & Crossing probability (in time-direction) \\
        $\mathcal C$ & Mutual information of two ancillas coupled at $i$ and $j$ & Probability $i$ and $j$ are in the same cluster \\ 
         $I_2(A,C)$ & Bipartite mutual information & $\max\{0, S(A)+S(C) - S(B) - S(D)$\}  \\
         $I_3(A,B,C)$ & Tripartite mutual information & $I_2(A,C) + S(B) + S(D) - S(A\cup B) - S(B \cup C) $ \\
        \hline \hline
    \end{tabular}
    \caption{In this table, regions $A$, $B$, $C$, and $D$ follow the geometry given in \cref{fig:subregions}, so that $A\cup B$ and $B \cup C$ are contiguous regions. $\mathcal C$ is what we refer to as the \emph{correlation function} and $i$ and $j$ are taken to be separated by $L/2$ throughout. Note that $I_2$ and $I_3$ are particular combinations of minimal cuts inspired to cancel divergent terms (as $L\rightarrow \infty$) present due to the geometry of the regions \cite{CasiniHuerta2004, KitaevPreskill2006}.}
    \label{tab:dictionary}
\end{table*}

To examine the entanglement properties of this model we consider the behavior of several quantities: the half-cut R\'enyi entropy $S_{L/2}$ and the entropy $S_a$ of an initially entangled ancilla as described above, as well as the bipartite and tripartite mutual informations $I_2$ and $I_3$, the purification time $t_\mathrm{pure}$, and the correlation $\mathcal{C}$ between two ancillas coupled into the system at equal time and $L/2$ sites apart; see \cref{tab:dictionary} for a quick reference on what these correspond to in different contexts. 

\begin{figure}
    \centering
    \includegraphics[width=0.25\columnwidth]{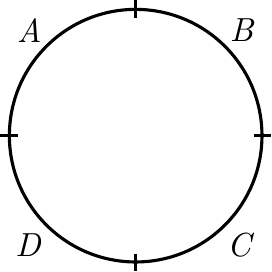}
    \caption{The division of the system into the four subregions labeled $A$, $B$, $C$, and $D$ considered in computing $I_2$ and $I_3$.}
    \label{fig:subregions}
\end{figure}

The mutual information is a measure of the delocalization of information, or, for our purposes here, the amount of entanglement between different subsystems. 
For the bi- and tripartite mutual informations we consider partitioning our system into 4 subsystems, each of size $L/4$, which we label $A$, $B$, $C$, and $D$, arranged as shown in \cref{fig:subregions}.
The bipartite mutual information $I_2$ is then expressed as 
\begin{equation} \label{eq:I2}
    I_2(A,C) = S(A) + S(C) - S(A\cup C),
\end{equation}
where $S(X)$ is the R\'enyi entropy from tracing out subsystem $X$, which for contiguous $X$ corresponds to a particular element of $\hat{d}$ as discussed in \cref{sec:minimalcutmodel}---$d_{ij}$ is the R\'enyi entropy for the subsystem consisting of qudits $i-1$ through $j$.
The last quantity in this expression, $S(A\cup C)$, is the R\'enyi entropy associated with a non-contiguous subsystem and so is not itself contained in $\hat{d}$, but can be constructed from its elements. 
It is most easily understood as the minimal cut needed to separate the domain(s) containing both $A$ and $C$ from the domain(s) containing $B$ and $D$, which can be made in two possible ways--either we cut out $A$ and $C$ or we cut out $B$ and $D$.
Where therefore want the minimum of $S(A)+S(C)$ and $S(B)+S(D)$, which are trivial to compute from $\hat{d}$.

The tripartite mutual information $I_3$ considers the same partitioning of the system as above, and can be expressed as
\begin{multline} \label{eq:I3}
    I_3(A,B,C) = I_2(A,B) + I_2(B,C) - I_2(B,A\cup C) \\
    = S(A) + S(B) + S(C) + S(A\cup B\cup C)\\
    - S(A\cup B) - S(B\cup C) - S(A\cup C).
\end{multline}
Since the regions $A$ and $B$ are adjacent, $S(A\cup B)$ is just the half-cut entropy calculated for that particular subsystem (similarly for $S(B\cup C)$), and the quantity $S(A\cup B\cup C)$ is equivalent to $S(D)$.
For the model in question, $I_3$ is always non-positive \cite{Zabalo2020}. 

The purification time is defined in terms of the ancilla entropy, since this is related to the system's purity---for a given circuit realization, $t_\mathrm{pure}$ is the first time step at which $S_a = 0$. 
In a volume-law phase this time is exponentially large in $L$, but since we cannot simulate a circuit for exponentially long time, we instead find the average purification time $\overline{t}_\mathrm{pure}$ in this phase to saturate to the time depth of the circuit, $t_\mathrm{max} = 2 L^2$. 
In area-law and disentangled phases the value we find is more representative of the actual time for the system to purify, though still lower than the true average value because of the rare circuit realizations that are still not pure by $t_\mathrm{max}$. 

Finally, we consider the correlation $\mathcal{C}$ between two ancillas locally entangled into the system at time $t = t_\mathrm{max}/2$ and separated by a distance $L/2$.
This can be precisely formulated as the mutual information $I_2$, defined above, between these two ancilla qubits.
In our classical model, however, it is more simply understood via the percolation picture of connected clusters---the correlation is $1$ if both ancillas are connected to sites in the same cluster and $0$ otherwise---and it is with this perspective that we calculate it.  
We first evolve the system for time $t_\mathrm{max}/2$ while tracking the connectivity matrix $\hat{c}$ as in \cref{sec:ancillamodel}, then couple the ancillas into the system with two vectors $a_1$ and $a_2$, initialized as 
\begin{equation}
    a_{1,i} = c_{1,i} \qquad a_{2,i} = c_{L/2+1,i},
\end{equation}
so that ancilla $1$ is coupled to site $1$ and therefore connected to all sites connected to site $1$, and similarly for ancilla $2$ with site $L/2+1$.
We then evolve the system for the remaining time, updating $a_{1,2}$ in the same way as $a$ in \cref{sec:ancillamodel}. 
Because the elements of $a_1$ track the connectivity of ancilla $1$ to the $L$ sites of the system, and similarly for $a_2$ with ancilla $2$, if $a_{1,L/2+1}=1$ (or equivalently $a_{2,1}=1$) at any point in this process then $\mathcal{C}=1$, and otherwise $\mathcal{C}=0$. 
All of these observables are listed in \cref{tab:dictionary} in terms of their notation and what they corresponds to in the quantum information and Potts model contexts.

In all cases we examine these quantities averaged over many realizations of our stochastic circuit protocol, which we denote with a bar, e.g., $\overline{S}_{L/2}$.
We use 1000 realizations for these averages all quantities except the correlation function, for which we use 10\,000 realizations. 
Additionally, we will consider a range of system sizes---$L=8, 12, 16, 20, 24, 36, 48, 60, 72, 84, 96, 108$ and $120$---with the ranges used different circumstances indicated on corresponding plots.  
Criticality only truly manifests in the thermodynamic limit $L\to\infty$, so we use the change in the behavior of observables with increasing $L$ to determine critical properties.
To locate the boundaries between different phases we evaluate $\overline{S}_a$, $\overline{S}_{L/2}$, $\overline{I}_2$, $\overline{I}_3$, and $\overline{t}_\mathrm{pure}$ along various slices through the phase diagram, for example varying $q$ while holding $p$ constant. 
After locating the phase boundaries, we further examine their critical properties using the time dependence of $\overline{S}_a$ and $\overline{S}_{L/2}$ and the form of the correlation $\overline{\mathcal{C}}$ at the critical points. 

\section{Results and Discussion}
\label{sec:results}

\subsection{Percolation Criticality} 
\label{sec:percolation}

\begin{figure*}[!ht]
    \centering
    \includegraphics[width=0.9\textwidth]{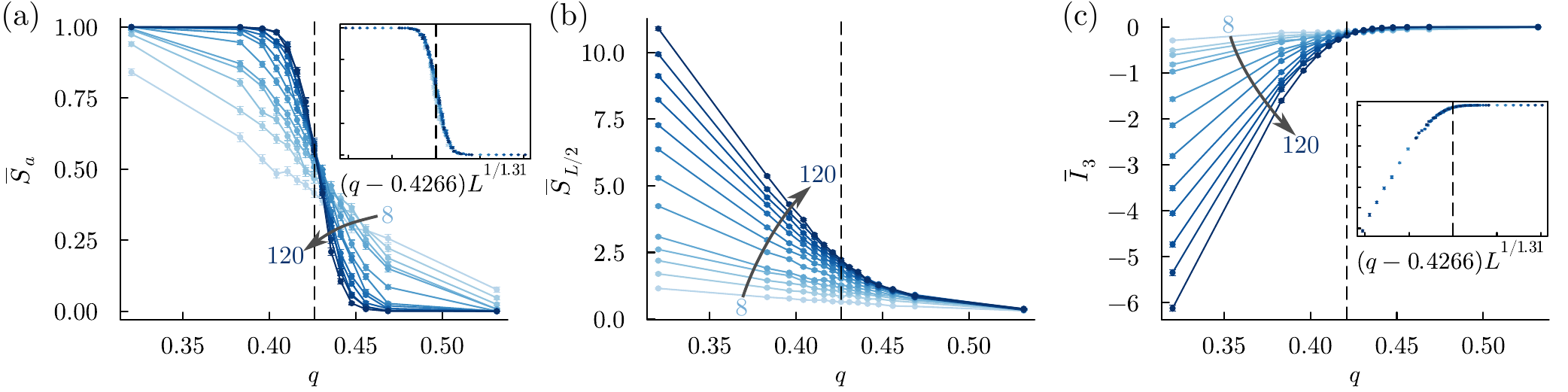}
    \caption{\textbf{Vertical slices:} The $q$-dependence of (a) the ancilla entropy $\overline{S}_a$, (b) the half-cut entropy $\overline{S}_{L/2}$, and (c) the tripartite mutual information $\overline{I}_3$ across the transition from volume-law to area-law entangled phases for the representative value $p=0.25$.
    The insets of (a) and (c) show collapse of $\overline{S}_a$ and $\overline{I}_3$ computed for different system sizes using $q_c = 0.4262$ and $\nu_\mathrm{perc} = 1.29$.
    Quantities are evaluated after $2L^2$ time steps for indicated range of system sizes. 
    }
    \label{fig:verticalcuts}
\end{figure*}

First, we identify and examine the critical properties of the phase boundary between volume- and area-law entangled phases by considering vertical slices through the phase space---varying $q$ for fixed values of $p$.
In \cref{fig:verticalcuts} we show $\overline{S}_a$, $\overline{S}_{L/2}$, and $\overline{I}_3$ as functions of $q$ for $p=0.25$, representative of the generic behavior along the phase boundary. 
For $p=0$ the circuit has a regular structure, and in addition to the algorithms described in \cref{sec:minimalcutmodel,sec:ancillamodel} we can determine $\overline{S}_{L/2}$ and $\overline{S}_a$ via a wetting algorithm as in Ref.~\cite{Skinner2019}.
The two methods are found to give the same results, and we confirm that the $p=0$ critical point can be understood via percolation in a spatially regular lattice in \cref{app:percolationlimit}. 
In \cref{fig:verticalcuts} we show $\overline{S}_a$, $\overline{S}_{L/2}$, and $\overline{I}_3$ along a vertical slice more representative of the generic behavior along the phase boundary, varying $q$ for fixed $p=0.25$.
The value of $q_c$ for each $p$, identifying the critical line, can be determined in several ways---from crossings in $\overline{S}_a$ and $\overline{I}_3$ for different values of $L$, and the point where the coefficient of the linear-in-$L$ behavior of $\overline{S}_{L/2}$ well within the volume-law phase extrapolates to 0.
The results from all three methods are in good agreement and are used to draw the phase diagram, \cref{fig:phasediagram}. 
The crossing of the ancilla entropy gives the most precise value, and two-parameter collapse of this data lets us additionally extract the correlation length critical exponent $\nu_\mathrm{perc}$ along the critical line. 
Because small systems are affected by the twist in the periodic boundary conditions, causing a drift in the crossing for small $L$, we consider systematically eliminating small $L$ to identify these critical properties, discussed in \cref{app:finitesize}.
Values of $q_c$ and $\nu_\mathrm{perc}$ obtained from this process are given in \cref{tab:perc_criticality}.
The values of $\nu_\mathrm{perc}$ are consistent with the percolation value $\nu=4/3$ for $p \leq 0.4$, but for larger $p$ the critical behavior is overwhelmed by the critical fan of the control transition at $p=1/2$, analyzed in \cref{sec:randomwalk} below.

For $p=0$ the circuit has a regular structure, and in addition to the analysis above we can determine $\overline{S}_{L/2}$ and $\overline{S}_a$ via a wetting algorithm as in Ref.~\cite{Skinner2019}.
The two methods are found to give the same results (see \cref{app:percolationlimit}), confirming that the $p=0$ critical point can be understood via percolation. 

To further characterize the critical behavior of this transition, we examine the time dependence of $\overline{S}_a$ and $\overline{S}_{L/2}$, and the correlation function $\overline{\mathcal{C}}$ along the phase boundary as determined from the ancilla entropy. 
The time-dependence of the half-cut entropy is expected to have a logarithmic form for early times due to conformal symmetry at the phase transition, just as the $L$ dependence at criticality is logarithmic for late times,
\begin{equation}\label{eq:Slog}
    \overline{S}_{L/2} \sim \begin{cases} \alpha_L \log(L), & T \gg L,\\
    \alpha_T \log(T), & T \ll L,
    \end{cases}
\end{equation} 
where $\alpha_{L,T}$ are universal coefficients that we can extract numerically.
The dynamical critical exponent $z$ relates spatial and temporal scaling, $T\sim L^z$, which gives the relation $\alpha_T = \alpha_L/z$. 
Similarly, the ancilla entropy collapses as a function of $T/L^z$, giving us two independent methods to obtain this exponent.
The $L$ dependence of the correlation function is expected to have power-law form $\overline{\mathcal{C}} \sim L^{-\eta}$.
In the percolation universality class, these quantities have the known values $z=1$, $\alpha_L=\alpha_T\approx0.54$, and $\eta=5/24=0.208\overline{33}$.
Because there is uncertainty in the value of the critical point $q_c$ for any given $p$, we evaluate these quantities not at just $q_c$, but also at $q_c\pm\sigma_{q_c}$ and use the results at these additional points to set uncertainties for our estimates of these quantities. 

\begin{figure*}[!ht]
    \centering
    \includegraphics[width=0.9\textwidth]{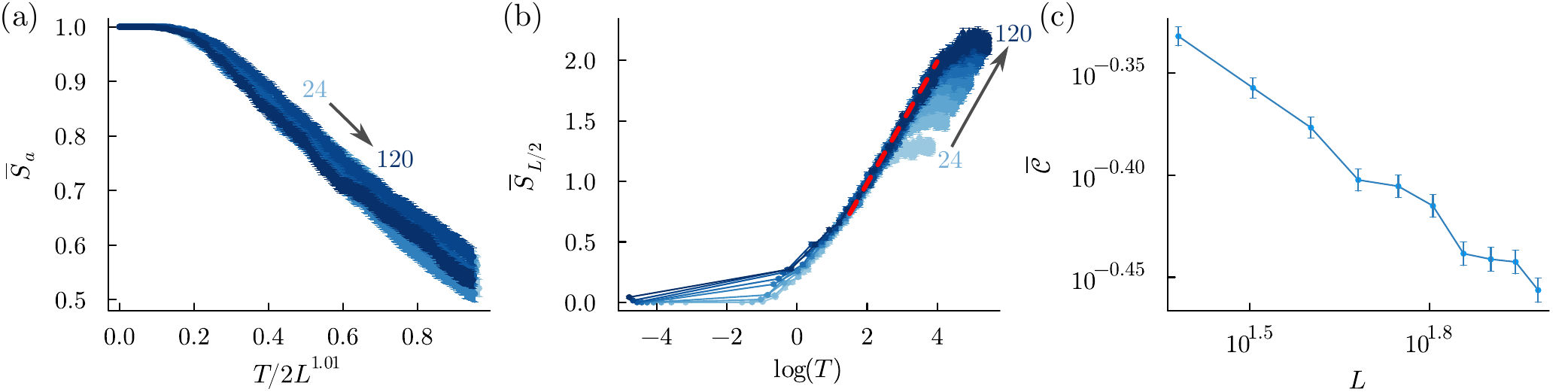}
    \caption{The time dependence of (a) the ancilla entropy $\overline{S}_a$ and (b) the half-cut entropy $\overline{S}_{L/2}$, both collapsed as functions of $T=t/L$, and (c) the correlation function $\overline{\mathcal{C}}$, all evaluated at the point along the volume-to-area-law transition at $p=0.25$.
    The dashed red line in (b) demonstrates the logarithmic growth of $\overline{S}_{L/2}$ before saturation at late time. 
    Note that (c) uses a log-log plot, showing the power-law behavior of $\overline{\mathcal{C}}$.
    }
    \label{fig:perctimedep}
\end{figure*}

The collapse of $\overline{S}_a$ and $\overline{S}_{L/2}$ as functions of time, and the behavior of the correlation function, are shown in \cref{fig:perctimedep} for the representative point at $p=0.25$ along the critical line.
We find that fitting the early-time behavior of the half-cut entropy to the logarithmic form \cref{eq:Slog} requires a rescaling of time~\cite{LeMaire2023}; if $t$ counts the time steps of the simulations then the time relevant for the dynamics of the entanglement at criticality is $T \propto t/L$.
With this rescaling in mind we collapse the ancilla entropy as a function of $T/L^{z_\mathrm{perc}}$, and we also fit the correlation $\overline{\mathcal{C}}$ to a power law form, with exponent $\eta$. 
The values of $z_\mathrm{perc}$ and $\eta$ thus obtained are given in \cref{tab:perc_criticality}, and show good agreement with percolation criticality along nearly the entire critical line.
The coefficient $\alpha_T$ characterizing the half-cut entropy dynamics, and therefore the values of $z_\mathrm{perc}$ determined from it, are much more sensitive to the uncertainty in the value of $q_c$ than the ancilla, but still yield values consistent with percolation criticality (see \cref{app:halfcuttime}). 

The need for this rescaling of time can be understood by considering how the sequential chaotic and control operations that build the circuit are correlated in space; the decimal point, determining where operations are applied, moves left or right through the system with each time step based on the value of $p$.
At long times there is an overall drift of this reference point with a velocity $v_p = 2p-1$, producing a circuit comprised of primarily gates ($p<1/2$) or measurements ($p>1/2$) along its trajectory.
For example, in \cref{fig:statmechschematic}(d) we see a chain of gates moving to the right. 
Starting at a fixed position $x$ in the system, it takes $L/\abs{v_p}$ time steps (on average) for the decimal point to travel completely around the system so that operations are applied near $x$ again, and after $t$ time steps the decimal point returns to its starting point $T = \abs{v_p} t/L$ times.
Therefore for $p<1/2$, for which there is an excess of chaotic steps each coming with a unitary gate, this $T$ can be interpreted as an effective circuit depth, explaining why it is the relevant time for entanglement dynamics.
We stress that this is not circuit depth as normally understood in quantum algorithms.

\begin{table*}
    \centering
    \setlength{\tabcolsep}{12pt}
    \begin{tabular}{c|cc|cc}
    \hline\hline
        $p$  & $q_c$ & $\nu_\mathrm{perc}$ & $z_\mathrm{perc}$ & $\eta$ \\
        \hline
        0.00 & 0.5012(20) & 1.38(18) & $0.99^{+0.07}_{-0.05}$ & $0.236^{+0.053}_{-0.045}$ \\
        0.05 & 0.4909(19) & 1.35(16) & $1.01^{+0.10}_{-0.05}$ & $0.224^{+0.068}_{-0.033}$ \\
        0.10 & 0.4790(16) & 1.30(15) & $1.00^{+0.08}_{-0.05}$ & $0.203^{+0.061}_{-0.011}$ \\
        0.15 & 0.4646(17) & 1.36(17) & $1.02^{+0.05}_{-0.05}$ & $0.207^{+0.026}_{-0.026}$ \\
        0.20 & 0.4488(17) & 1.28(14) & $1.00^{+0.05}_{-0.06}$ & $0.231^{+0.059}_{-0.043}$ \\
        0.25 & 0.4266(19) & 1.31(15) & $1.01^{+0.05}_{-0.06}$ & $0.205^{+0.048}_{-0.050}$ \\
        0.30 & 0.4004(22) & 1.34(16) & $1.01^{+0.10}_{-0.06}$ & $0.205^{+0.059}_{-0.032}$ \\
        0.35 & 0.3623(25) & 1.27(17) & $1.01^{+0.07}_{-0.05}$ & $0.187^{+0.046}_{-0.027}$ \\
        0.40 & 0.3046(32) & 1.34(17) & $1.07^{+0.08}_{-0.10}$ & $0.128^{+0.047}_{-0.038}$ \\
        0.45* & 0.221(4) & 1.20(15) & $1.09^{+0.09}_{-0.10}$ & $0.120^{+0.058}_{-0.036}$ \\
        0.48* & 0.153(7) & 0.97(9) & $1.000^{+0.004}_{-0.004}$ & $0.390^{+0.103}_{-0.113}$ \\
        \hline\hline
    \end{tabular}
    \caption{The values of the critical measurement probability $q_c$, correlation length exponent $\nu_\mathrm{perc}$, dynamical exponent $z_\mathrm{perc}$, and correlation exponent $\eta$ along the volume-law to area-law transition determined by data collapse of the ancilla entropy $S_a$.
    Because the crossing of the ancilla entropy drifts for small systems, to obtain $q_c$ and $\nu_\mathrm{perc}$ values we have considered only $L\geq 36$.
    (See \cref{app:finitesize} for values obtained with other choices of this minimum size.)
    The percolation universality class has $\nu = 4/3 = 1.\overline{33}$, $z=1$, and $\eta = 5/24 = 0.208\overline{33}$, and we find good agreement with these values along most of this critical line;
    The largest values of $p$, marked with an asterisk, show the most significant deviation due to the critical fan of the random walk at $p=1/2$.}
    \label{tab:perc_criticality}
\end{table*}

\subsection{Criticality Along \texorpdfstring{$p=1/2$}{Random Walk} Line} \label{sec:randomwalk}

\begin{figure*}[!ht]
    \centering
    \includegraphics[width=0.9\textwidth]{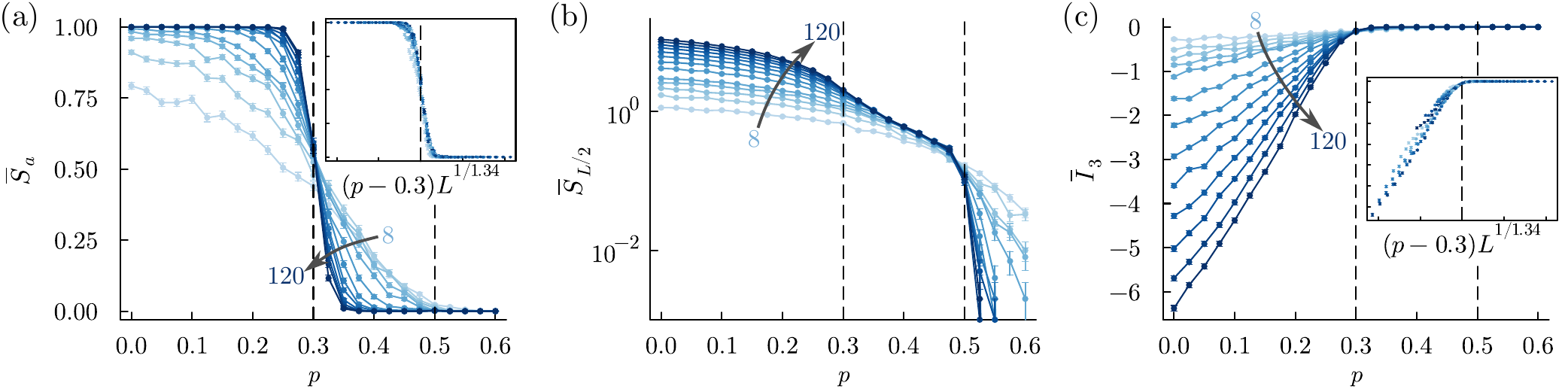}
    \caption{\textbf{Horizontal slices:} The $p$-dependence of (a) the ancilla entropy $\overline{S}_a$, (b) the half-cut entropy $\overline{S}_{L/2}$, and (c) the tripartite mutual information $\overline{I}_3$ at fixed value $q=0.4$ to identify the transitions from volume-law to area-law entanglement, and from area-law to disentangled.
    The insets of (a) and (c) show collapse of $\overline{S}_a$ and $\overline{I}_3$ computed for different system sizes using $p_c = 0.3$ and $\nu_\mathrm{perc} = 1.35$.
    Quantities are evaluated after $2L^2$ time steps for indicated range of system sizes. 
    }
    \label{fig:horizontalcuts}
\end{figure*}

As demonstrated in \cref{fig:verticalcuts}, for $p<1/2$ a nontrivial amount of entanglement remains in the system at long times on---we find either volume- or area-law-entangled phases.
On the other hand, for $p>1/2$ there is an excess of control steps, so in typical realization of the system almost all sites are measured at late times and we are left with a disentangled state.
At $p=1/2$ the same number of chaotic and control steps are applied on average; 
the bias velocity $v_p$ of the decimal point vanishes, and it instead executes a random walk through the system, so we expect the disentangling transition to display the universal behavior associated with this random walk. 
We investigate the nature of this transition by evaluating our chosen observables on horizontal slices through the phase diagram---sweeping $p$ for fixed values of $q$.
The case of $q=0.4$, representative of the generic behavior for all $q \in (0.0,0.5)$, is shown in \cref{fig:horizontalcuts}.

As expected from the results of \cref{sec:percolation}, for $q=0.4$ the transition from volume- to area-law behavior occurs at $p\approx0.3$.
The presence of small but non-zero values of $\overline{S}_a$ in the range $0.3<p<0.5$ for the smaller values of $L$ considered is indicative of area-law behavior; the system does not purify in $O(1)$ time as it does in the disentangled phase, and realizations that do not purify by the final time $t=2L^2$, more common in smaller systems, yield non-zero average ancilla entropy.
The clearest indication of two distinct transitions, however, is the behavior of the half-cut entropy, \cref{fig:horizontalcuts}(b).
Here we $\overline{S}_{L/2} \sim L$ behavior see collapse to an $L$-independent $O(1)$ value in the range $0.3\lesssim p<1/2$, followed by a sharp drop to very small values for $p>1/2$.
Though not show, the lower critical point is found to move to smaller $p$ with increasing $q$ until it reaches $p=0$ at $q=1/2$, and for $q>1/2$ the system has only a single transition from area-law to disentangled at $p=1/2$.

\begin{figure}[!ht]
    \centering
    \includegraphics[width=0.7\columnwidth]{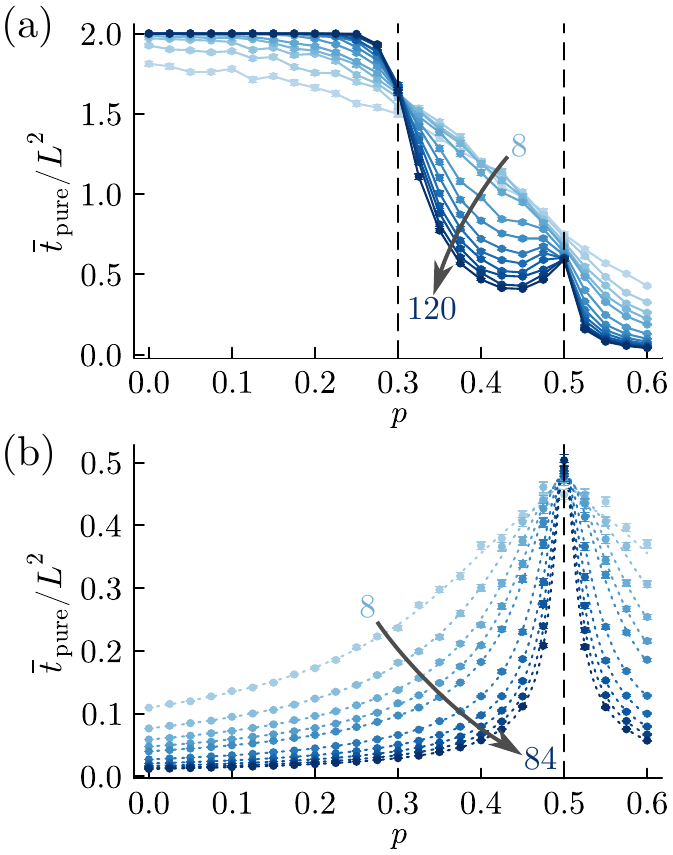}
    \caption{The $p$-dependence of the time for the ancilla to purify scaled by the square of the system size for (a) $q=0.4$ and (b) $q=1$.
    In (a) we see clear features at two values, $p\approx 0.3$ and $p=1/2$, corresponding the the transitions between volume- and area-law entangled phases, and between area-law and disentangled phases. 
    In (b) points (with error bars) are obtained from ancilla simulations, and dashed lines are from modeling the random walk of the decimal point---the two agree very precisely.
    We see a single feature at $p=1/2$ consistent with the argument in the main text.
    Purification time data is averaged over 1000 circuit realizations, and for the random walker over 1.6 million realizations.}
    \label{fig:tpure}
\end{figure}

We also see indications of these two transitions in the average purification time $\overline{t}_\mathrm{pure}$, shown scaled by the square of the system size in \cref{fig:tpure}(a) for $q=0.4$.
As already noted above, in the volume-law phase the purification time is exponential in $L$, much larger than the $2L^2$ time steps that we run our simulations for, so $\overline{t}_\mathrm{pure}$ saturates to this value.
However, even with this caveat we see that rescaling the average purification time by $L^2$ causes the curves for different $L$ to cross or pinch together for the same two points, identified via the behavior of $\overline{S}_{L/2}$ as indicative of the two transitions.

\begin{figure}
    \centering
    \includegraphics[width=\columnwidth]{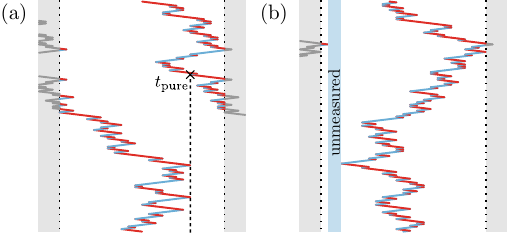}
    \caption{(a) A random walk that wraps the system, giving a finite purification time. (b) A random walk that does not wrap the system, leaving some sites unmeasured, so the system is not purified.}
    \label{fig:randomwalk}
\end{figure}

We can understand this behavior and furthermore how both random-walk and percolation behavior emerge by returning to the motion of the decimal point. 
As previously discussed, for $p<1/2$ this point drifts with nonzero velocity $v_p=2p-1$, resulting in an effective circuit depth $T=\abs{v_p}t/L$---the total number of time steps divided by the average time to wrap the system---and we needed to consider the time dependence of $\overline{S}_a$ and $\overline{S}_{L/2}$ in terms of $T$ to find consistent critical properties.
At $p=1/2$ the velocity vanishes and the decimal point instead executes an unbiased random walk, so the time to wrap the system changes parametrically to $O(L^2)$ and the effective depth is $T \sim t/L^2$.
Reasoning in terms of $T$ now helps us determine the purification time based on what we know from percolation and normal MIPTs.
At the percolation transition the ancilla purifies at a \emph{depth} of $T \sim O(L)$; this corresponds to a \emph{time} $t_\mathrm{pure} \sim O(L^2)$. 
In the area-law phase, however, the ancilla purifies at a depth $T \sim O(1)$ which gives $t_\mathrm{pure} \sim O(L)$ for $p<1/2$, but for $p \approx 1/2$ instead gives $t_\mathrm{pure} \sim O(L^2)$.
All of this is confirmed by our plot of the purification time for $q=0.4$ in \cref{fig:tpure}(a).
Furthermore, this reasoning also applies for purification in the extreme $q=1$ case, which now corresponds to the time for the random walk to traverse all points on the cylinder.
For example, notice in \cref{fig:randomwalk}(a) that the random walker wraps the cylinder and in \cref{fig:randomwalk}(b) it does not; these correspond to a disentangled ancilla and entangled ancilla, respectively.
This can be separately simulated and the two are found to match, as shown in \cref{fig:tpure}(b).

To see the interplay between this transition and the nearby percolation phase boundary for small $q$ we focus closer to $p=1/2$ and evaluate our chosen quantities at a time $t=L^2/2$, closer to the average purification time of the random walk so that the disentangling transition is more readily observed. 
This change has the desired effect on the ancilla and half-cut entropies, however the crossing in the tripartite mutual information $\overline{I}_3$ only tracks the percolation phase boundary. 
Collapsing the obtained values of the ancilla and half-cut entropies for $p>0.5$ we extract both the critical point $p_c$ as a function of $q$ as well as the correlation length critical exponent $\nu_\mathrm{RW}$, given in \cref{tab:RW_criticality}. 
From both entropies we find $p_c\approx 0.5$ and correlation length exponents $\nu_\mathrm{RW}\approx 1$ for all values of $q$ considered, demonstrating that this random walk behavior is unaffected by the nearby percolation transition. 

\begin{figure*}[!ht]
    \centering
    \includegraphics[width=0.9\textwidth]{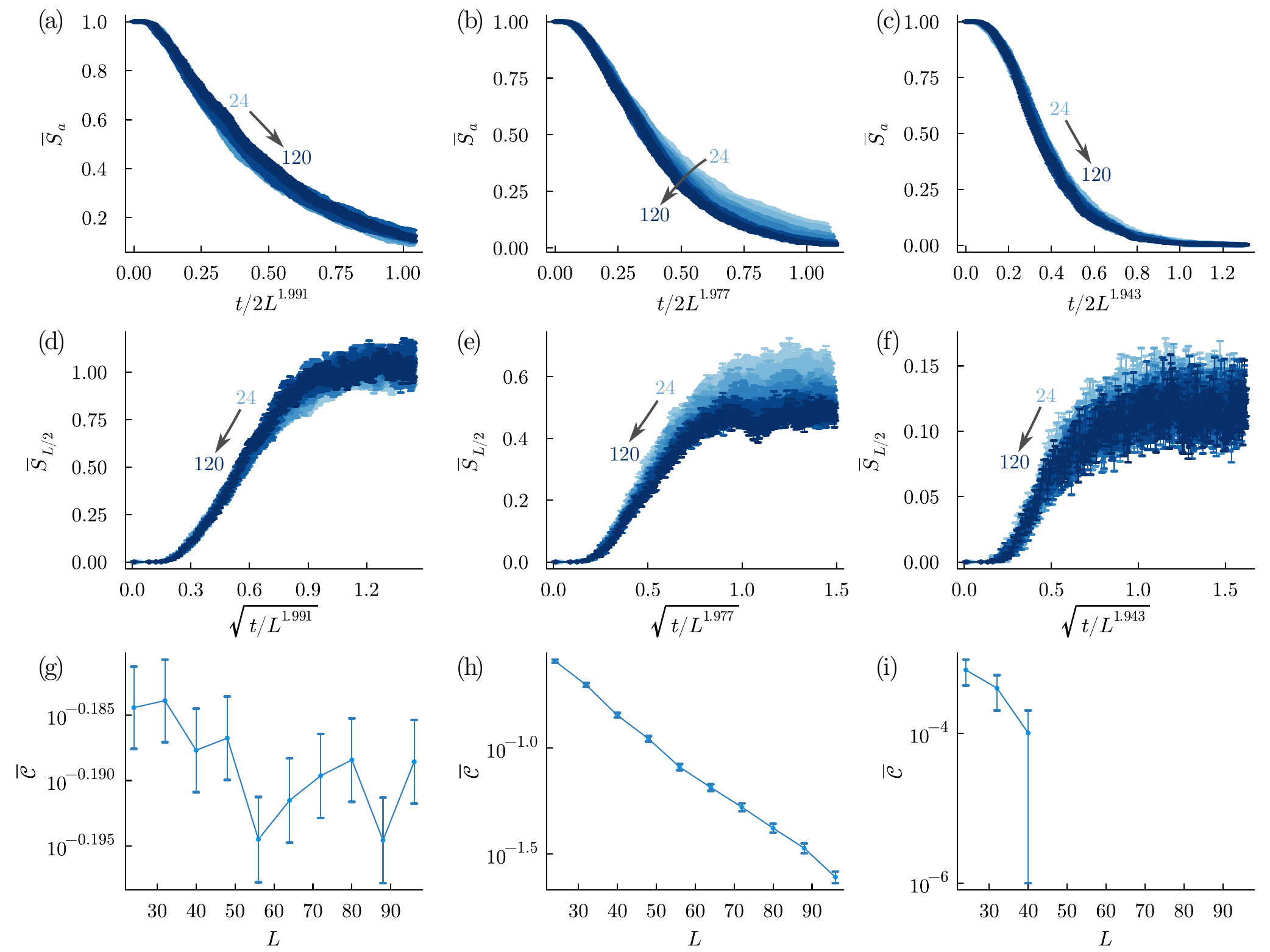}
    \caption{The time dependence of the ancilla (a,b,c) and half-cut (d,e,f) entropies, and the correlation function (g,h,i) along the line $p=1/2$ for three representative values of $q$: $q=0$ for (a,d,g), $q=0.1$ for (b,e,h), and $q=0.4$ for (c,f,i).}
    \label{fig:05timedata}
\end{figure*}

We also examine the time dependence of $\overline{S}_a$ and $\overline{S}_{L/2}$ and the correlation function $\overline{\mathcal{C}}$ for points along this critical line, shown in \cref{fig:05timedata}.
The ancilla and half-cut entropies for different $L$ collapse by rescaling time as $t/L^{z_\mathrm{RW}}$ with $z_\mathrm{RW}\approx2$ for all $q$.
The values of this exponent extracted from the time-dependence of the ancilla entropy are given in \cref{tab:RW_criticality}.
We expect the value that $\overline{S}_{L/2}$ saturates to at late times to be a universal quantity, but fluctuations from the volume-to-area-law transition very close to this vertical line for small $q$ strongly affect small systems.
This lead to a deviation for small $L$, seen most clearly in \cref{fig:05timedata}(e), and making it difficult to estimate $z_\mathrm{RW}$ from $\overline{S}_{L/2}$. 
The correlation function $\overline{\mathcal{C}}$ is found to have exponential form along $p=1/2$;
because this quantity is taken directly from percolation theory (it is the probability that the two sites spatially separated by $L/2$ are in the same cluster), it should only behave critically, i.e. as a power-law, at the percolation transition. 
At $p=1/2$ (and $q \neq 0$) only small connected clusters remain in a typical realization of the circuit, so we should expect only exponentially small correlations at distances $\sim L$. 

\begin{table*}
    \centering
    \setlength{\tabcolsep}{12pt}
    \begin{tabular}{c|cc|cc|c}
    \hline\hline
        $q$ & $p_c$ (half-cut) & $\nu_\mathrm{RW}$ (half-cut) & $p_c$ (ancilla) & $\nu_\mathrm{RW}$ (ancilla) & $ z_\mathrm{RW}$ (ancilla)\\
        \hline
        0.0 & 0.5018(21) & 0.88(20)& 0.4996(27) & 1.02(16) & 1.991(9)\\
        0.1 & 0.4959(29) & 1.14(24)& 0.4983(27) & 1.06(17) & 1.977(16)\\
        0.2 & 0.5007(32) & 0.85(26)& 0.4973(31) & 1.12(17) & 1.943(9)\\
        0.3 & 0.5001(34) & 0.88(30)& 0.497(4) & 1.04(24) & 1.978(6)\\
        0.4 & 0.499(4) & 0.97(26)& 0.500(6) & 1.01(32) & 1.943(10)\\
        \hline\hline
    \end{tabular}
    \caption{The values of $p_c$ and $\nu_\mathrm{RW}$ that collapse the half-cut $\overline{S}_{L/2}$ and ancilla $\overline{S}_a$ entropies evaluated at a final time $L^2/2$ for several fixed values of $q$ across $p=1/2$, and the values of $z_\mathrm{RW}$ that collapse $\overline{S}_a$ as a function of time (up to time $L^2/2$) when evaluated at $p=1/2$.}
    \label{tab:RW_criticality}
\end{table*}

\subsection{Properties of the volume-law phase} 
\label{sec:volumephase}

We have already noted some features of the volume-law phase with the quantities that identified the volume-law phase boundary.
The behavior of the tripartite mutual information demonstrates the presence of long-range entanglement in the system, purification takes exponentially long in $L$ as evidenced by the ancilla entropy, and the half-cut entanglement entropy scales linearly with $L$.
Now looking to the bipartite mutual information $I_2$, we can examine the sub-leading $L$-dependence of the R\'enyi entropies as well. 

\begin{figure}[!ht]
    \centering
    \includegraphics[width=0.7\columnwidth]{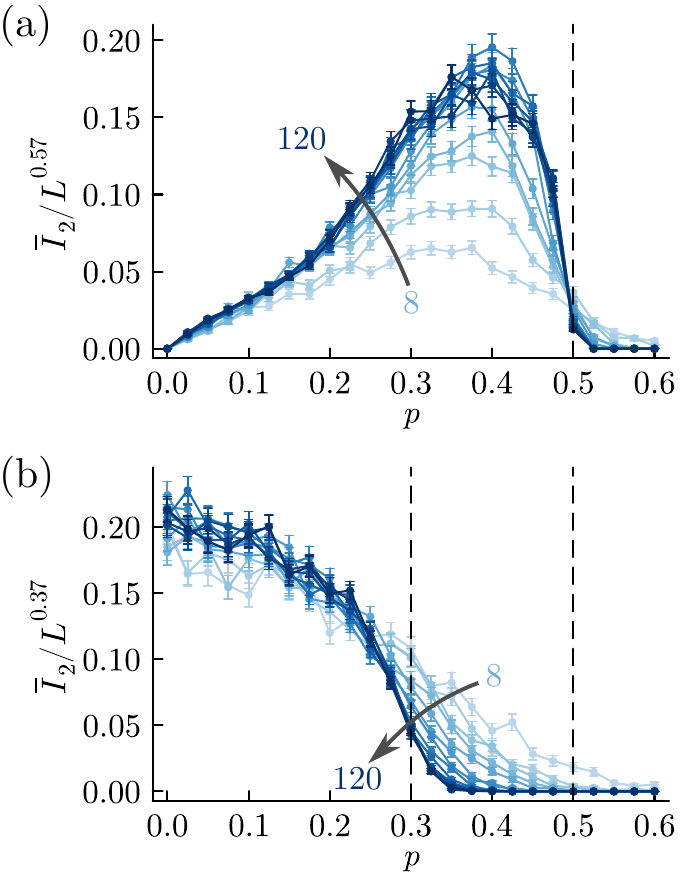}
    \caption{The bipartite mutual information $\overline{I}_2$ as a function of $p$ for (a) $q=0$ and (b) $q=0.4$, scaled by $L^\beta$ to collapse for large $L$.
    For $q=0$ we find collapse with $\beta = 0.57(10)$, and for $q=0.4$ we find $\beta = 0.37(10)$.
    Vertical dotted lines mark the entanglement transitions present for each $q$.
    }
    \label{fig:I2scaling}
\end{figure}

Within the volume-law phase of systems exhibiting standard MIPTs, the entanglement of a region $A$ of size $\ell$ grows like $\bar{S}(A) = a \ell + b \ell^{\beta} + \cdots$ at long times, with universal exponent $\beta = 1/3$ obtained from Kardar-Parisi-Zhang (KPZ) universality \cite{PotterVasseur2022a,FisherVijay2023}.
This quantity is captured by $I_2(A,C) \sim L^{\beta}$, and indeed when $q = 0.4$ we observe a scaling consistent with this, shown in \cref{fig:I2scaling}(b), and extract $\beta = 0.37(10)$. 
This ceases to be the case for the more correlated dynamics when $q=0$, and indeed we start to get a new, larger exponent $\beta = 0.57(10)$, as seen in \cref{fig:I2scaling}(a). 
For $q=0$, ``cuts'' in the entanglement are strictly performed during the control step as one can see in the schematic \cref{fig:statmechschematic}(d); due to this, even a heuristic mapping to a Poisson growth process for a height model should not apply, and indeed we find that subleading entanglement growth exceeds KPZ.
While we do not have a full theory for this subleading growth, we speculate that it is unstable for finite $q$, flowing to KPZ for $q>0$.

\section{Conclusion and Outlook} 
\label{sec:conclusion}

Here we have developed a classical statistical mechanics model that provides insight into the entanglement properties of a quantum analog of the $d$-adic R\'enyi map under stochastic control.
By including both stochastic control operations and additional random projective measurements this model yields two distinct classical critical behaviors---bond percolation in 2d and a random walk in 1d---each characterizing a specific entanglement transition of the original quantum model. 
We therefore have three distinct classical models giving us insight into the entanglement and controllability of our single quantum model:
the volume-to-area-law entanglement transition is dual to a model of bond percolation, the disentangling transition can be understood as arising from random-walk physics, and the controllability transition (concurrent with disentangling) is captured by the original $d$-adic map with stochastic control.

This unifies these three models from previous work in a model that has each as particular limits;
the MIPT of a brickwork circuit with random measurements emerges here for $p=0$, and the disentangling/controllability transition characterized by a random walk studied in Ref.~\cite{Iadecola2023} is obtained here for $q=0$---two boundaries of the phase diagram \cref{fig:phasediagram}.
The model developed here is able to flesh out the details of the 2d phase space in between, continuously interpolating between these two distinct models and showing how their unique critical properties interact within the entanglement and purification structure.
Specifically, along the volume-law phase boundary we see percolation criticality become overwhelmed by the random walk as we approach the transition at $p=1/2$. 

It is important to note that the details of our results depend on an implicit assumption regarding the control operation, namely that it involves only \emph{local} operations.
Indeed, as described in \cref{sec:Renyimodel}, control onto the state $\ket{00\dots0}$ can be accomplished via a reset operation built from solely local measurements and feedback.
As conceptualized, this feedback can be implemented with single-qudit operations and has no impact on the statistical mechanics model.
The Potts model we obtain lives on a random lattice due to the stochastic nature of its generation, but contains only local connectivity between its degrees of freedom.  
If we were to consider a different target state, however, control would in general require a more complicated feedback involving non-local operations, such as an adder circuit~\cite{Iadecola2023}, which will introduce non-local spreading of entanglement and therefore must result in non-local connectivity within the statistical mechanics mapping.
Understanding how to incorporate such nonlocal control into the statistical mechanics model is an interesting direction for future work.

Another open question concerns the effect of finite-$d$ corrections to the infinite-$d$ results obtained here.
For example, it would be interesting to determine whether the coincidence of the entanglement and control transitions on the $q=0$ line persists at finite $d$.
Indeed, in Ref.~\cite{SierantTurkeshi2023} it was found that such coincidence occurs for long-range control protocols and is generically absent in short-range ones.
The control operation considered in this work is strictly local, but the large-$d$ limit may also affect the interplay of control and entanglement.
This further emphasizes the need to develop an improved understanding of the necessary and sufficient conditions for control and entanglement transitions to coincide.

\begin{acknowledgements}
We would like to thank Michael Buchhold, Haining Pan, and Jed Pixley for helpful discussions and comments. 
This work was supported in part by the National Science Foundation under Grants No.~DMR-2238895 (C.~L. \& J.~H.~W.) and No.~DMR-2143635 (T.~I.)
\end{acknowledgements}
\newpage
\appendix

\section{Percolation Limit}
\label{app:percolationlimit}

\begin{figure}[!t]
    \centering
    \includegraphics[width=0.8\columnwidth]{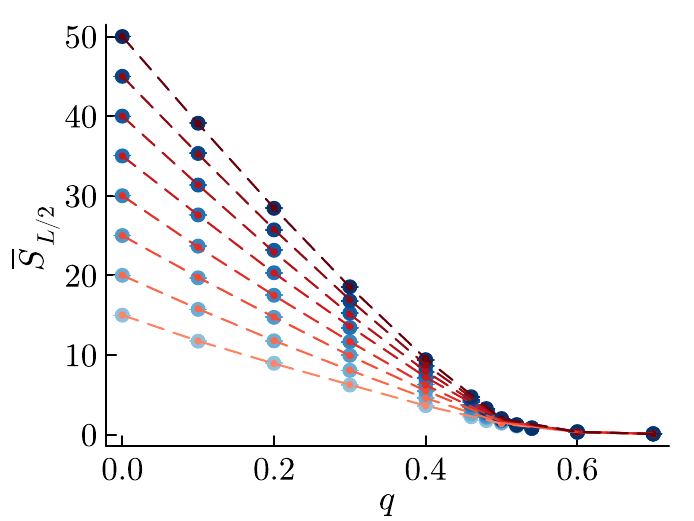}
    \caption{Comparison of the half-cut entropy determined as minimal cuts via algorithm described in \cref{sec:minimalcutmodel} for $p=0$ (blue points) and the wetting algorithm on the tilted square lattice with twisted boundary conditions (red points).
    The two are in very good agreement.
    We use system sizes $L=30, 40, 50, 60, 70, 80, 90$ and $100$, and averages are taken over 1000 circuit realizations.}
    \label{fig:wetting}
\end{figure}

In the $p = 0$ limit, we recover percolation with a particular twist in the boundary conditions as illustrated in \cref{fig:twisted}.
In order to both verify this claim and validate the stat-mech algorithm used throughout the main text to evaluate $\overline{S}_{L/2}$, we implement a wetting algorithm \cite{Skinner2019} with these boundary conditions.
We see in \cref{fig:wetting} that the resulting entropies match what was found for $p=0$ in the main text. 

\section{Finite-size effects and collapse}
\label{app:finitesize}

The nature of our circuit protocol induces a small shift in the periodic boundary condition of our system. 
Small systems are most sensitive to the effects of this shift, which is realized as a drift in the crossing of the ancilla entropy $\overline{S}_a$ for small $L$, which vanishes in the thermodynamic limit.
To determine the value of the percolation critical line via the ancilla crossing, we therefore consider systematically removing the smallest systems until the obtained values of $q_c$ and $\nu_\mathrm{perc}$ for each $p$ becomes sufficiently insensitive to further removals.
In \cref{tab:perc_criticality} we quote values for $q_c$ and $\nu_\mathrm{perc}$ keeping systems of size $L=36$ and above.
Here in \cref{tab:perc_qcnu} we show how the effect of changing this small system cutoff, either by keeping or excluding an additional system size. 
We see that keeping only $L\geq48$ yields values that differ from those obtained with $L\geq36$ by less than the error in either case, while the same cannot be said for the values obtained with $L\geq24$. 
Therefore, we use the $L\geq36$ values in our analysis. 

\begin{table*}
    \centering
    \setlength{\tabcolsep}{12pt}
    \begin{tabular}{c|ccc|ccc}
    \hline\hline
          & \multicolumn{3}{c|}{$q_c$} & \multicolumn{3}{c}{$\nu_\mathrm{perc}$} \\
        $p$ & $L \geq 24$ & $L \geq 36$ & $L \geq 48$ & $L \geq 24$ & $L \geq 36$ & $L \geq 48$ \\
        \hline
        0.00 & 0.5012(15) & 0.5012(20) & 0.5007(18) & 1.38(14) & 1.38(18) & 1.32(19) \\
        0.05 & 0.4908(15) & 0.4909(19) & 0.4906(21) & 1.37(14) & 1.35(16) & 1.34(20) \\
        0.10 & 0.4796(16) & 0.4790(16) & 0.4789(20) & 1.31(13) & 1.30(15) & 1.32(17) \\
        0.15 & 0.4651(13) & 0.4646(17) & 0.4641(18) & 1.22(10) & 1.36(17) & 1.31(19) \\
        0.20 & 0.4488(14) & 0.4488(17) & 0.4488(22) & 1.27(11) & 1.28(14) & 1.29(17) \\
        0.25 & 0.4277(17) & 0.4266(19) & 0.4262(21) & 1.29(12) & 1.31(15) & 1.29(16) \\
        0.30 & 0.3999(19) & 0.4004(22) & 0.4001(27) & 1.36(14) & 1.34(16) & 1.35(19) \\
        0.35 & 0.3628(24) & 0.3623(25) & 0.3617(31) & 1.38(16) & 1.27(17) & 1.30(21) \\
        0.40 & 0.3073(26) & 0.3046(32) & 0.3053(35) & 1.24(12) & 1.34(17) & 1.32(19) \\
        0.45* & 0.2251(33) & 0.221(4) & 0.218(4) & 1.16(12) & 1.20(15) & 1.13(13) \\
        0.48* & 0.163(6) & 0.153(7) & 0.149(8) & 0.87(8) & 0.97(9) & 0.9(1) \\
        \hline\hline
    \end{tabular}
    \caption{The values of the critical measurement probability $q_c$ and correlation length exponent $\nu_\mathrm{perc}$ along the volume-law to area-law transition determined by data collapse of the ancilla entropy $S_a$.
    The three columns for each quantity correspond to the range of system sizes used in this process---the drift in the crossing of the ancilla entropy for small systems, a finite size effect, motivates discarding smaller system sizes. 
    The largest values of $p$, marked with an asterisk, show the most significant deviation from percolation universality due to the critical fan of the random walk at $p=1/2$.}
    \label{tab:perc_qcnu}
\end{table*}

For the transition at $p=1/2$, characterized by the universality of a random walk, we do not find the same sensitivity to the periodic boundary condition, as evidenced by the insensitivity of $p_c$ and $\nu_\mathrm{RW}$ determined from both $\overline{S}_a$ and $\overline{S}_{L/2}$, shown in \cref{tab:RW_ancilla_full,tab:RW_halfcut_full}, to this same procedure. 

\begin{table*}
    \centering
    \setlength{\tabcolsep}{12pt}
    \begin{tabular}{c|ccc|ccc}
    \hline\hline
          & \multicolumn{3}{c|}{$p_c$ (ancilla)} & \multicolumn{3}{c}{$\nu_\mathrm{RW}$ (ancilla)} \\
        $q$ & $L \geq 8$ & $L \geq 12$ & $L \geq 16$ & $L \geq 8$ & $L \geq 12$ & $L \geq 16$ \\
        \hline
        0.0 & 0.4996(27) & 0.4994(25) & 0.4998(24) & 1.02(16) & 1.04(15) & 1.01(14) \\
        0.1 & 0.4983(27) & 0.4984(27) & 0.4993(25) & 1.06(17) & 1.05(17) & 1.02(16) \\
        0.2 & 0.4973(31) & 0.4979(32). & 0.498(3) & 1.12(17) & 1.04(20) & 1.10(17) \\
        0.3 & 0.497(4) & 0.498(4) & 0.498(4) & 1.04(24) & 1.10(21) & 1.10(21) \\
        0.4 & 0.500(6) & 0.497(4) & 0.500(4) & 1.01(32) & 1.14(21) & 1.01(24) \\
        \hline\hline
    \end{tabular}
    \caption{The values of $p_c$ and $\nu_\mathrm{RW}$ that collapse the ancilla entropy $\overline{S}_a$ evaluated at a final time $L^2/2$ for several fixed values of $q$ across $p=1/2$ and several choices for the smallest system size retained.
    We see that these values are not greatly affected by systematic removal of small systems, and that we obtain values consistent with random walk universality, $p_c = 1/2$ and $\nu_\mathrm{RW} = 1$, in all cases; unlike for the percolation critical line, there is no drift in the ancilla entropy crossing.}
    \label{tab:RW_ancilla_full}
\end{table*}

\begin{table*}
    \centering
    \setlength{\tabcolsep}{12pt}
    \begin{tabular}{c|ccc|ccc}
    \hline\hline
          & \multicolumn{3}{c|}{$p_c$ (half-cut)} & \multicolumn{3}{c}{$\nu_\mathrm{RW}$ (half-cut)} \\
        $q$ & $L \geq 8$ & $L \geq 12$ & $L \geq 16$ & $L \geq 8$ & $L \geq 12$ & $L \geq 16$ \\
        \hline
        0.0 & 0.5018(21) & 0.5016(24) & 0.5004(20) & 0.88(20) & 0.91(22) & 0.97(19) \\
        0.1 & 0.4959(29) & 0.4976(28) & 0.4977(28) & 1.14(24) & 1.02(25) & 1.09(24) \\
        0.2 & 0.5007(32) & 0.5003(29). & 0.5009(29) & 0.85(26) & 0.89(25) & 0.84(28) \\
        0.3 & 0.5001(34) & 0.4972(31) & 0.5009(32) & 0.88(30) & 1.10(18) & 0.84(33) \\
        0.4 & 0.499(4) & 0.498(4) & 0.497(4) & 0.97(26) & 1.1(4) & 1.1(4) \\
        \hline\hline
    \end{tabular}
    \caption{The values of $p_c$ and $\nu_\mathrm{RW}$ that collapse the half-cut entropy $S_{L/2}$ evaluated at a final time $L^2/2$ for several fixed values of $q$ across $p=1/2$.
    We see that these values are not greatly affected by systematic removal of small systems, and that we obtain values consistent with random walk universality, $p_c = 1/2$ and $\nu_\mathrm{RW} = 1$, in all cases.}
    \label{tab:RW_halfcut_full}
\end{table*}

\begin{widetext}
    
\section{Time-dependence of \texorpdfstring{$\overline{S}_{L/2}$}{half-cut entropy} along percolation transition}
\label{app:halfcuttime}

Here we discuss the time dependence of the half-cut entropy $\overline{S}_{L/2}$ along the percolation critical line.  
For our chosen points along this critical line the coefficient $\alpha_T$ is extracted from a fit of the time dependence of $\overline{S}_{L/2}$ for $L=120$ to the expected logarithmic behavior at early times.
Examining the value the half-cut entropy saturates to at late-times for different $L$, we can likewise determine the coefficient $\alpha_L$, then determine the dynamic critical exponent $z_\mathrm{perc}$ as the ratio of these $\alpha's$.
The values obtained for $\alpha_L$, $\alpha_T$, and $z_\mathrm{perc} = \alpha_L/\alpha_T$ are given in \cref{tab:perc_z}.
We find that $\alpha_L$ and $z_\mathrm{perc}$ are consistent with the universal values $\alpha \approx 0.54$ and $z=1$ due to fairly large uncertainties, though $\alpha_T$ tends to be too small along most of the phase boundary. 

\begin{table*}
    \centering
    \setlength{\tabcolsep}{12pt}
    \begin{tabular}{c|ccc}
    \hline\hline
        $p$ & $\alpha_L$ & $\alpha_T$ & $z_\mathrm{perc}=\alpha_L/\alpha_T$ \\
        \hline
        0.00 & $0.516^{+0.077}_{-0.044}$ & $0.499^{+0.026}_{-0.117}$ & $1.03(29)$ \\
        0.05 & $0.527^{+0.069}_{-0.057}$ & $0.471^{+0.026}_{-0.096}$ & $1.12(27)$ \\
        0.10 & $0.536^{+0.050}_{-0.040}$ & $0.446^{+0.067}_{-0.030}$ & $1.20(21)$ \\
        0.15 & $0.562^{+0.054}_{-0.054}$ & $0.477^{+0.042}_{-0.049}$ & $1.18(17)$ \\
        0.20 & $0.514^{+0.051}_{-0.037}$ & $0.459^{+0.046}_{-0.030}$ & $1.12(16)$ \\
        0.25 & $0.547^{+0.053}_{-0.044}$ & $0.502^{+0.024}_{-0.078}$ & $1.09(20)$ \\
        0.30 & $0.516^{+0.059}_{-0.043}$ & $0.447^{+0.024}_{-0.025}$ & $1.15(15)$ \\
        0.35 & $0.519^{+0.046}_{-0.061}$ & $0.476^{+0.023}_{-0.062}$ & $1.09(19)$ \\
        0.40 & $0.563^{+0.046}_{-0.081}$ & $0.522^{+0.026}_{-0.024}$ & $1.08(16)$ \\
        0.45* & $0.468^{+0.070}_{-0.046}$ & $0.546^{+0.019}_{-0.065}$ & $0.86(16)$ \\
        0.48* & $0.279^{+0.063}_{-0.036}$ & $0.529^{+0.015}_{-0.024}$ & $0.53(12)$ \\
    \hline\hline
    \end{tabular}
    \caption{The values of the coefficients of the logarithmic behavior of the half-cut entropy for $T\gg L$ ($\alpha_L$) and $T\ll L$ ($\alpha_T$), and the dynamical critical exponent $z_\mathrm{perc}$ determined as their ratio for points along the volume-law to area-law transition. 
    The percolation universality class has $\alpha_L = \alpha_T \approx 0.54$, and $z=1$.
    We find values for $z_\mathrm{perc}$ and $\alpha_L$ largely consistent with this universality for most of this phase boundary, though $\alpha_T$ tends to be too small.
    The largest values of $p$, marked with an asterisk, show the most significant deviation from percolation universality due to the critical fan of the random walk at $p=1/2$.}
    \label{tab:perc_z}
\end{table*}

\end{widetext}

\bibliography{references}

\end{document}